%% file: main.tex
\documentclass[oneside,12pt]{book}

\usepackage{fancyvrb}
\usepackage[utf8]{inputenc}

\usepackage[arabic,english,french]{babel}
\usepackage{graphicx}

\usepackage[a4paper,total={6in, 8in}]{geometry}
\usepackage{fancyhdr}
\usepackage[titletoc,title]{appendix}
\usepackage{float}
\usepackage{caption}

\fancypagestyle{plain}{%
	\fancyhf{}%
	\fancyfoot[C]{\thepage}%
}

\usepackage{amsmath}
\usepackage{algorithm}
\usepackage{algorithmic}

\usepackage{xcolor}
\usepackage[most]{tcolorbox}
\usepackage{listings}

\definecolor{white}{rgb}{1,1,1}
\definecolor{mygreen}{rgb}{0.00,0.44,0.13}
\definecolor{light_gray}{rgb}{0.97,0.97,0.97}
\definecolor{mykey}{rgb}{0.117,0.403,0.713}
\definecolor{comment}{rgb}{0.38,0.63,0.69}
\definecolor{red}{rgb}{1,0,0}

\tcbuselibrary{listings}
\newlength\inwd
\usepackage{url}

\newcounter{ipythcntr}
\renewcommand{\theipythcntr}{\texttt{[\arabic{ipythcntr}]}}

\newtcblisting{pyin}[1][]{%
	sharp corners,
	enlarge left by=\inwd,
	width=\linewidth-\inwd,
	enhanced,
	boxrule=0pt,
	colback=light_gray,
	listing only,
	top=0pt,
	bottom=0pt,
	overlay={
		\node[
		anchor=north east,
		text width=\inwd,
		font=\footnotesize\ttfamily\color{mykey},
		inner ysep=2mm,
		inner xsep=0pt,
		outer sep=0pt
		] 
		at (frame.north west)
		{\refstepcounter{ipythcntr}\label{#1}In \theipythcntr:};
	}
	listing engine=listing,
	listing options={
		aboveskip=0.5pt,
		belowskip=0.5pt,
		basicstyle=\footnotesize\ttfamily,
		language=Python,
		keywordstyle=\color{mygreen},
		showstringspaces=false,
		stringstyle=\color{red},
		commentstyle=\color{comment},
		morekeywords={True,False},
		numberstyle=\tiny\color{mygreen}
	},
}
\newtcblisting{pyprint}{
	sharp corners,
	enlarge left by=\inwd,
	width=\linewidth-\inwd,
	enhanced,
	boxrule=0pt,
	colback=white,
	listing only,
	top=0pt,
	bottom=0pt,
	overlay={
		\node[
		anchor=north east,
		text width=\inwd,
		font=\footnotesize\ttfamily\color{mykey},
		inner ysep=2mm,
		inner xsep=0pt,
		outer sep=0pt
		] 
		at (frame.north west)
		{};
	}
	listing engine=listing,
	listing options={
		aboveskip=1pt,
		belowskip=1pt,
		basicstyle=\footnotesize\ttfamily,
		language=Python,
		keywordstyle=\color{mykey},
		showstringspaces=false,
		stringstyle=\color{mygreen},
		numberstyle=\tiny\color{mygreen}
	},
}
\newtcblisting{pyout}[1][\theipythcntr]{
	sharp corners,
	enlarge left by=\inwd,
	width=\linewidth-\inwd,
	enhanced,
	boxrule=0pt,
	colback=white,
	listing only,
	top=0pt,
	bottom=0pt,
	overlay={
		\node[
		anchor=north east,
		text width=\inwd,
		font=\footnotesize\ttfamily\color{red},
		inner ysep=2mm,
		inner xsep=0pt,
		outer sep=0pt
		] 
		at (frame.north west)
		{\setcounter{ipythcntr}{\value{ipythcntr}}Out#1:};
	}
	listing engine=listing,
	listing options={
		aboveskip=1pt,
		belowskip=1pt,
		basicstyle=\footnotesize\ttfamily,
		language=Python,
		keywordstyle=\color{mykey},
		showstringspaces=false,
		stringstyle=\color{mygreen}
	},
}

\let\mylistof\listof
\renewcommand\listof[2]{\mylistof{algorithm}{Liste des algorithmes}}
\makeatletter
\providecommand*{\toclevel@algorithm}{0}
\makeatother

\addto\captionsfrench{}

\usepackage{setspace}

\begin{document}

\input{Chapitres/PageDeGarde}

\input{Chapitres/Remerciement}

\input{Chapitres/Dedicaces}



\pagenumbering{roman}
\begin{spacing}{1.4}
\tableofcontents
\listoffigures
\listofalgorithms
\end{spacing}
\cleardoublepage
\renewcommand{\labelitemi}{$\bullet$}
\renewcommand{\labelitemii}{$\ast$}
\renewcommand{\labelitemiii}{$\diamond$}
\renewcommand{\labelitemiv}{$\ast$}

\input{Chapitres/Introduction}
\input{Chapitres/Chapitre-01}

\input{Chapitres/Chapitre-02}
\input{Chapitres/Chapitre-03}

\input{Chapitres/Chapitre-04}

\nocite{ref1,ref2,ref3,ref4,ref5,ref6,ref7,ref8,ref9,ref10,ref11,ref12,ref13,ref14,ref15,ref16,ref17,ref18,ref19,ref20}

\input{Chapitres/Conclusion}

\bibliographystyle{apalike}
\addcontentsline{toc}{chapter}{Références bibliographiques}
\bibliography{Chapitres/Bibliographie}

\input{Chapitres/Annexe}

\input{Chapitres/Resume}

\end{document}

%% file: Chapitres/PageDeGarde.tex
\sloppy
\begin{titlepage}
\begin{center}
{\bf République Algérienne Démocratique et Populaire
\\ Ministère de L’Enseignement Supérieur et de La Recherche Scientifique\\
Université Ferhat Abbas Sétif-1}\\
\begin{center}
\includegraphics{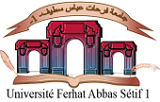}
\end{center}
{\bf Faculté des Sciences} \\ {\bf Département d'Informatique}\\
\vspace{0.1cm}
\Large{\emph{{{\it {Mémoire de fin de cycle }}}}}\\
\vspace{0.1cm}
\normalsize
\begin{center}
En vue d’obtention du diplôme de
Master académique \\ en Informatique
\end{center}
\normalsize\textbf{Option: \\}Ingénierie des Données et Technologie Web

\Huge\textbf{Thème}\\
\noindent\rule{\textwidth}{0.9mm}
\Large{{\textbf{Quantification de la propagation des rumeurs sur les e-réseaux-sociaux}}}
\noindent\rule{\textwidth}{0.9mm}
\end{center}
\vspace{0.5cm}

\begin{tabular}[t]{@{}l} 
	
	\Large\textbf{\textbf{Présenté par :}}\vspace{0.5cm}\\

	\large $M^{lle} $ OUNOUGHI Chahinez\\
	\vspace{0.3cm}
\end{tabular}
\hfill
\begin{tabular}[t]{l@{}}
	\Large\textbf{\textbf{Encadré par : }}\vspace{0.5cm}\\
	\large $M^{r} Dr.$ \textsl{BENAOUDA } Abdelhafid
	\vspace{0.3cm}
\end{tabular}
\vspace{-0.3cm}\\

\begin{center}
	\large Année universitaire 2017/2018
\end{center}
\end{titlepage}

%% file: Chapitres/Remerciement.tex
\chapter*{Remerciements}
	
\thispagestyle{empty}

\textit{En tout premier lieu, je remercie le bon Dieu Allah, tout puissant, de m’avoir donné la force pour survivre, ainsi que l’audace pour dépasser toutes les difficultés et de réaliser ce travail.}

\vspace{0.3cm}
\textit{La réalisation de ce mémoire a été possible grâce au concours de plusieurs personnes à qui je voudrais témoigner toute ma reconnaissance.}

\vspace{0.3cm}
\textit{\textbf{À NOTRE MAITRE, ENCADREUR DE MÉMOIRE :}}

\vspace{0.3cm}
\textit{\textbf{DOCTEUR BENAOUDA ABDELHAFID}, je suis très honoré de vous avoir comme encadreur de notre mémoire. Je vous remercie pour la gentillesse et la spontanéité avec lesquelles vous avez bien voulu diriger ce travail. J'ai eu le grand plaisir de travailler sous votre direction, et j'ai trouvé auprès de vous le conseiller et le guide qui me a reçu en toute circonstance avec sympathie, sourire et bienveillance. Veuillez, cher Maître, trouver dans ce modeste travail l'expression de ma haute considération, de ma sincère reconnaissance et de mon profond respect. 
}

\vspace{0.3cm}
\textit{\textbf{À MA FAMILLE :}}

\vspace{0.3cm}
\textit{Un très grand MERCI à toute ma famille qui m’a gratifié de son amour et fourni les motivations qui ont permis l’aboutissement de mes études. Je leur adresse toute ma gratitude du fond du mon cœur.} 

\vspace{0.3cm}
\textit{\textbf{À MES AMIS :}}

\vspace{0.3cm}
\textit{Pour tous mes amis qui m’ont apporté leur soutien moral pendant ces années d’études, je les en remercie sincèrement.}

\vspace{0.3cm}

%% file: Chapitres/Dedicaces.tex
\chapter*{Dédicaces}
	
\thispagestyle{empty}

\textit{\textbf{À MES CHERS PARENTS :}}

\vspace{0.3cm}
\textit{Aucune dédicace ne saurait exprimer mon respect, mon amour éternel et ma considération pour les sacrifices que vous avez consenti pour mon instruction et mon bien être. Je vous remercie pour tout le soutien et l’amour que vous me portez depuis mon enfance et j’espère que votre bénédiction m’accompagne toujours.Que ce modeste travail soit l’exaucement de vos vœux tant formulés, le fruit de vos innombrables sacrifices, bien que je ne vous en acquitterai jamais assez. Puisse Dieu, le Très Haut, vous accorder santé, bonheur et longue vie et faire en sorte que jamais je ne vous déçoive.}

\vspace{0.3cm}
\textit{\textbf{À MES CHERS ET ADORABLES SŒURS ET FRÈRE :}}

\vspace{0.3cm}
\textit{\textbf{Meriem}, la douce, au cœur si grand, \textbf{Nihad} la généreuse, \textbf{Douaa} l’aimable, \textbf{Mohammed Houssam EDDine} mon petit frère que j’adore.  En témoignage de mon affection fraternelle, de ma profonde tendresse et reconnaissance, je vous souhaite une vie pleine de bonheur et de succès et que Dieu, le tout puissant, vous protége et vous garde. }

\vspace{0.3cm}
\textit{\textbf{À MES AMIS DE TOUJOURS :}}  

\vspace{0.3cm}
\textit{\textbf{ASMAA}, \textbf{KENZA} ET \textbf{ISMAHEN} ; En souvenir de notre sincère et profonde amitié et des moments agréables que nous avons passés ensemble. Veuillez trouver dans ce travail l’expression de mon respect le plus profond et mon affection la plus sincère. 
}

\clearpage

%% file: Chapitres/Introduction.tex
\chapter*{Introduction générale}
\pagenumbering{arabic}
\markboth{\textit{Introduction générale}}{}
\addcontentsline{toc}{chapter}{Introduction générale}

Au cours de la dernière décennie, Internet est devenu un acteur majeur en tant que source d'information. Une étude du Pew Research Center a identifié Internet comme la ressource la plus importante pour les news des personnes de moins de 30 ans aux États-Unis et la deuxième plus importante après la télévision \cite{PewCenter}.
 
\vspace{0.3cm}
Plus récemment, l'émergence et l'augmentation de la popularité des réseaux sociaux et des services de réseautage tels que \textit{Twitter} et \textit{Facebook} ont grandement affecté les reportages d'actualité et les paysages journalistiques. Alors que les réseaux sociaux sont surtout utilisés pour le bavardage (chat) quotidien, ils sont devenu comme source d'information; Cela est particulièrement vrai pour les situations de news de rupture (breaking-news), où les gens ont besoin de mises à jour rapides sur le développement d'événements en temps réel.

\vspace{0.3cm}
Comme \cite{Kwak.2010} ont montré que plus de 85\% de tous les sujets d'actualité sur Twitter sont des news. En outre, l'omniprésence, l'accessibilité, la rapidité et la facilité d'utilisation des réseaux sociaux ont fait des sources inestimables d'informations de première main. Cependant, les mêmes facteurs qui font des réseaux sociaux une excellente ressource de la diffusion de news de rupture (breaking-news), conjugué au manque relatif de supervision de ces services, font des réseaux sociaux un terreau fertile pour la création et la diffusion d'informations non étayées et non vérifiées sur les événements qui se produisent dans le monde. 

\newpage

Ce changement sans précédent des médias traditionnels, où il existe une distinction claire entre les journalistes et les consommateurs de news; ces derniers sont fournis dans les réseaux sociaux en ligne par la foule qui sont devenus aussi des journalistes.

\vspace{0.3cm}
À la différence, un journaliste est tenu à respecter une charte déontologie alors qu'un utilisateur lambda, activant sur les réseaux sociaux peut émettre n'importe quelle information sans que ce dernier soit inquiété.  Cet état de fait a présenté de nombreux défis pour divers secteurs de la société tels que les journalistes, services d'urgence et consommateurs de news. Les journalistes doivent maintenant rivaliser avec des millions de personnes en ligne pour obtenir des informations de dernière minute. Souvent, cela amène les journalistes à ne pas trouver un équilibre entre le besoin d'être premier et le besoin d'être correct, ce qui entraîne un nombre croissant de sources d'information rapportant des informations non fondées dans la précipitation pour être les premières. Les services d'urgence doivent faire face aux conséquences et aux retombées des rumeurs et des chasses aux sorcières sur les réseaux sociaux. Enfin, les consommateurs ont la lourde tâche de trier les messages afin de séparer les messages justifiés et dignes de confiance des rumeurs et des hypothèses injustifiées.

\vspace{0.3cm}
C'est dans ce cadre que nous voulons apporter notre contribution, à savoir, suivre la trace d'une information sur un réseau social. Cette trace peut avoir des profondeurs différentes selon l'importance ou l'impact de cette information, que nous appelons rumeur, en attendant sa confirmation. 

\vspace{0.3cm}
Le but de ce mémoire est justement de proposer une contribution académique afin de prédire et quantifier la profondeur de la propagation des rumeurs sur un réseau social.

\vspace{0.2cm}
Nos résultats pourront être exploités par des officiels ou des spécialistes en sciences humaines afin d'étudier et de comparer différents impacts d'une rumeur par rapport à d'autres et proposer des solutions afin de limiter la mauvaise diffusion des informations et particulièrement ce qu'on appelle actuellement les \textit{fake-news}. 

Notre contribution dans ce projet de fin d'études de master en informatique option Ingénierie de Données et Technologies de Web (IDTW)  est détaillée dans ce mémoire. Ce dernier est composé de quatre chapitres qui sont organisés comme suit : 
\vspace{0.2cm}
\begin{itemize}
	\item Le premier chapitre intitulé « \textbf{Généralités sur les réseaux sociaux} », c'est un tour d'horizon sur les réseaux sociaux, leurs types et caractéristiques ainsi que la notion des rumeurs.
	
	\vspace{0.3cm}
	\item Le deuxième chapitre nommé « \textbf{État de l’art} », présente une revue des travaux récents faits dans le monde académique sur la modélisation de la propagation d'information en général puis spécialement des rumeurs, en résumant avec des critiques concernant ces modèles.
	
	\vspace{0.3cm}
	\item Le troisième chapitre « \textbf{Une approche pour la quantification de la profondeur d'une rumeur} », la proposition de notre propre contribution, qui est une approche pour pouvoir prédire la diffusion de la rumeur dans les réseaux sociaux basant sur la notion de similarité.
	
	\vspace{0.3cm}
	\item Le dernier chapitre présente les différentes phases suivies lors de notre déploiement de l'approche proposée afin de mettre en œuvre, valider et interpréter les résultats obtenus par la solution proposée.
	
	\vspace{0.3cm}
\end{itemize}

Enfin, Ce mémoire se termine par une conclusion et une vision sur nos travaux futurs ainsi qu'un annexe présentant quelques caractéristiques techniques utilisées lors du déploiement et de l'implémentation de notre projet.  

%% file: Chapitres/Chapitre-01.tex
\chapter{Généralités sur les réseaux sociaux}

\section{Introduction}

Au cours des dernières années, les gens obtiennent de plus en plus rapidement de l'information grâce au développement de la technologie Internet. Les réseaux sociaux en tant que nouvelles plates-formes de soutien à la diffusion de l'information et à l'établissement des relations sociales deviennent  progressivement un pivot important de la vie des gens.

\vspace{0.3cm}
Dans ce chapitre, nous présentons d'abord un aperçu des réseaux sociaux, en décrivant la définition et leurs objectifs, puis les mécanismes et les politiques fournis par la plupart des réseaux sociaux. Ensuite, nous citons le problème des rumeurs dans ces réseaux.

\section{Définition et Objectifs}

Nous définissons les sites de réseaux sociaux comme des services Web permettant aux individus de :

\vspace{0.3cm}
\begin{itemize}
	\item  Construire un profil public ou semi-public dans un système délimité.
	\vspace{0.3cm}
	\item Créer des liens explicites vers d'autres utilisateurs avec lesquels, ils partagent une connexion.
	\vspace{0.3cm}
	\item Naviguer sur le réseau social en parcourant les liens et profils d'autres utilisateurs dans le système.
\end{itemize}
\vspace{0.3cm}

Les réseaux sociaux en ligne remplissent plusieurs objectifs, mais trois rôles principaux sont communs à tous les sites. 

\vspace{0.2cm}
Premièrement, les réseaux sociaux en ligne sont utilisés pour maintenir et renforcer les liens sociaux existants, ou pour créer de nouveaux liens sociaux.

\vspace{0.2cm}
Deuxièmement, les réseaux sociaux en ligne sont utilisés par chaque membre pour télécharger son propre contenu. Notons que le contenu partagé varie souvent d'un site à l'autre.

\vspace{0.2cm}
Troisièmement, les réseaux sociaux en ligne sont utilisés pour trouver de nouveaux contenus intéressants en filtrant, recommandant et organisant le contenu téléchargé par les utilisateurs.

\section{Historique}

Maintenant un bref historique des réseaux sociaux en ligne. Le site \textbf{Classmates.com} est considéré comme le premier site Web permettant aux utilisateurs de se connecter à d'autres. Il a commencé en \textbf{1995} en tant que site permettant aux utilisateurs de se reconnecter avec leurs anciens camarades de classe et compte actuellement plus de 70 millions d'utilisateurs enregistrés. Cependant, \textbf{Classmates.com} permettait aux utilisateurs de se lier les uns aux autres uniquement via les écoles qu'ils avaient fréquentées. En \textbf{1997}, le site \textbf{SixDegrees.com} a été créé, premier site du réseau social permettant aux utilisateurs de créer des liens directement avec d'autres utilisateurs.

\vspace{0.3cm}
Les réseaux sociaux en ligne ont commencé à gagner en popularité à mesure que de plus en plus d'utilisateurs se connectaient à Internet. Au début des années \textbf{2000}, un certain nombre de sites à vocation générale de trouver des amis, ont été établis; et le plus notable étant \textbf{Friendster}\footnote{http://www.friendster.com}.

\textbf{Friendster} se concentrait sur le fait de permettre à des amis d'amis de se rencontrer. \textbf{CyWorld\footnote{http://www.cyworld.com}}, \textbf{Ryze\footnote{https://www.ryze.com}} et \textbf{LinkedIn\footnote{https://www.linkedin.com}} sont d'autres sites similaires créés au cours de la même période.

\vspace{0.3cm}
En \textbf{2003}, \textbf{MySpace}\footnote{https://myspace.com/} a été créé comme une alternative à \textbf{Friendster} et les autres. \textbf{MySpace} est un réseau social dédié à la musique et aux artistes. Entièrement redesigné en 2013 après sa gloire passée, il s’est aujourd’hui recentré sur une cible plus restreinte.

\vspace{0.3cm}
\textbf{Facebook}\footnote{https://www.facebook.com}était créé au début de \textbf{2004} par Mark Zuckerberg sous la forme d'un réseau social destiné aux étudiants d'Harvard. Selon sa large propagation, le site est devenu grand public en \textbf{2006}. Il permet la découverte de nouveaux contenus, le suivi de la vie de vos proches, le chat et le partage des photos et des vidéos auprès de vos amis et de vos proches. Au début de \textbf{2018}, \textbf{Facebook} classé comme le plus large réseau social au monde avec \textbf{2,20} milliards d’utilisateurs actifs par mois \cite{fbinvestor}.

\vspace{0.3cm}
\textbf{Twitter\footnote{https://twitter.com}} a été créé par Jack Dorsey, Evan Williams, Biz Stone et Noah Glass, et lancé en 2006. Twitter est un site du réseau social dit de \textbf{«microblogging»} qui permet de communiquer sous la forme de messages courts ne dépassant pas 140 caractères appelés \textbf{« tweets »}. Les tweets peuvent contenir des URL sous forme raccourcie, des images, des émoticônes, des gifs animés et des vidéos.En avril \textbf{2018}, \textbf{Twitter} était classé le \textbf{17 \textsuperscript{ième}} réseau social au monde avec \textbf{336 millions} d’utilisateurs actifs par mois \cite{twitterq1}.

\vspace{-0.4cm}
\section{Types des réseaux sociaux}

Avec l'augmentation de la popularité des réseaux sociaux en ligne, de nombreux autres types de sites ont commencé à inclure des fonctionnalités de réseaux sociaux.

\begin{enumerate}
	\item \textbf{Sites de Connexions sociales}\\
	C'est  sites de masse sont des réseaux de personnes connectés par des systèmes d’amis et de fans. Voici une liste des sites Web les plus utilisés pour créer des liens sociaux en ligne :
	\begin{itemize}
		\item \textbf{Facebook} 
		Incontestablement l'outil de réseaux sociaux le plus populaire, il permet aux utilisateurs de créer des liens et de partager des informations avec les personnes et les organisations avec lesquelles ils choisissent d'interagir en ligne.
		
		\item \textbf{Twitter} 
		Partagez vos pensées et restez en contact avec les autres via ce réseau d'information en temps réel.
		
		\item \textbf{Google+\footnote{https://plus.google.com}} 
		Ce nouveau venu sur le marché des connexions sociales est conçu pour permettre aux utilisateurs de créer des cercles de contacts avec lesquels ils peuvent interagir et qui sont intégrés à d'autres produits Google.
		
		\item \textbf{MySpace} 
		Bien qu'il ait d'abord commencé comme un site général de médias sociaux, MySpace a évolué pour se concentrer sur le divertissement social, fournissant un lieu pour les connexions sociales liées aux films, aux jeux musicaux et plus encore.
		
	\end{itemize}

	\item \textbf{Sites de partage de contenu multimédia}\\
	Ces sites qui permettent la publication de contenus générés (vidéos, photos, .. etc) par les utilisateurs. Voici quelques-uns des sites les plus populaires pour le partage multimédia :
	
	\begin{itemize}
		\item \textbf{YouTube\footnote{https://www.youtube.com}} 
		Plate-forme de médias sociaux permettant aux utilisateurs de partager et d'afficher du contenu vidéo.
		
		\item \textbf{Flickr\footnote{https://www.flickr.com}} 
		Ce site offre une option puissante pour la gestion des photographies numériques en ligne, ainsi que pour les partage avec d'autres personnes.
		
	\end{itemize}
	
	\vspace{1cm}
	\item \textbf{Sites professionnels}\\
	Les réseaux sociaux professionnels sont conçus pour offrir des opportunités de croissance liées à la carrière. Certains de ces types de réseaux constituent un forum général pour les professionnels, tandis que d'autres se concentrent sur des professions ou des intérêts spécifiques. Quelques exemples de réseaux sociaux professionnels sont listés ci-dessous :
	
	\begin{itemize}
		
		\item \textbf{LinkedIn} 
		Les participants ont la possibilité d'établir des relations en établissant des liens et en rejoignant des groupes pertinents.
		
		\item \textbf{Classroom2.0\footnote{http://www.classroom20.com}} 
		Réseau spécialement conçu pour aider les enseignants à se connecter, partager et s'entraider avec des sujets spécifiques à la profession.
		
	\end{itemize}
	
	\item \textbf{Sites académiques}\\
	Les chercheurs universitaires qui souhaitent partager leurs recherches et examiner les résultats obtenus par leurs collègues peuvent trouver que le réseautage social propre à l'établissement est très utile. Quelques-unes des communautés en ligne les plus populaires pour les universitaires sont :
	
	\begin{itemize}
		
		\item \textbf{Academia.edu} 
		Les utilisateurs de ce réseau social universitaire peuvent partager leurs propres recherches, ainsi que suivre des recherches soumises par d'autres.
		
		\item \textbf{Connotea Collaborative Research\footnote{http://www.connotea.org}} 
		Ressource en ligne pour les scientifiques, les chercheurs et les praticiens cliniques pour trouver, organiser et partager des informations utiles.
	
	\end{itemize}
	
	\item \textbf{Sites éducatifs}\\
	Les réseaux éducatifs sont l'endroit où de nombreux étudiants vont collaborer avec d'autres étudiants sur des projets académiques, et de mener des recherches pour l'école, ou d'interagir avec des professeurs et des enseignants via des blogs et des forums en classe. Quelques exemples de tels réseaux sociaux éducatifs sont énumérés ci-dessous :
	
	\begin{itemize}
		
		\item \textbf{The Student Room\footnote{https://www.thestudentroom.co.uk}}
		Communauté estudiantine basée au Royaume-Uni(UK) avec un forum modéré et des ressources utiles liées à l'école.
		
		\item \textbf{The Math Forum\footnote{http://mathforum.org/students}}
		Un grand réseau éducatif conçu pour connecter les étudiants ayant un intérêt pour les mathématiques, ce site offre des possibilités d'interaction pour les étudiants par groupe d'âge.
	\end{itemize}
	
\end{enumerate}

\section{Cartographie des réseaux sociaux}

Une cartographie des réseaux sociaux éditée en Janvier 2018, montrant les sites de réseautage social les plus populaires par pays, selon \textit{Alexa\footnote{https://www.alexa.com}} et \textit{SimilarWeb\footnote{https://www.similarweb.com}}.

	\begin{center}
		\includegraphics[scale=0.4]{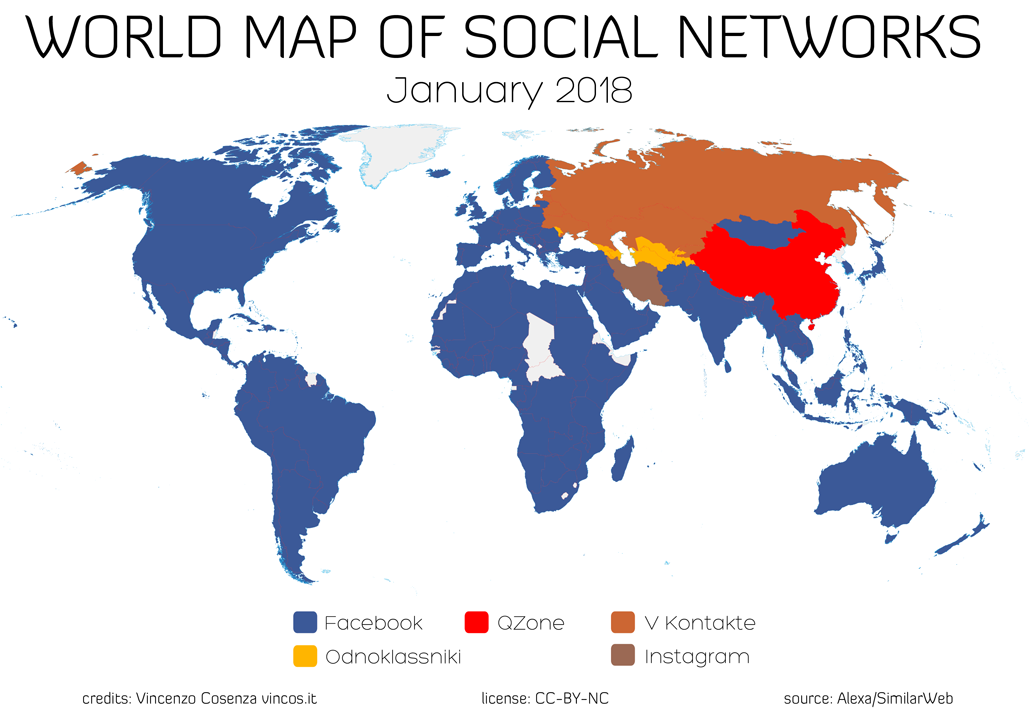}\\
		\captionof{figure}{Carte mondiale des réseaux sociaux.}
	\end{center}

\textbf{Facebook} semble plus fort que jamais. Maintenant, il est le premier réseau social dans 152 des 167 pays analysés (91\% de la planète).

\newpage
 Mark Zuckerberg n'a que trois antagonistes: \textbf{VK (VKontakte)\footnote{https://vk.com}} et \textbf{Odnoklassniki\footnote{https://ok.ru}} (appartenant au même groupe \textit{Mail.ru}) dans les territoires \textit{Russes}, et \textbf{QZone\footnote{https://qzone.qq.com}} en \textit{Chine}. En \textit{Iran}, suite à la censure de l'Etat envers Facebook, il y a une chance pour \textbf{Instagram\footnote{https://www.instagram.com}}.

\section{Mécanismes et politiques}

Un bref aperçu des mécanismes et des politiques que fournissent la plupart des réseaux sociaux en ligne.

\vspace{0.3cm}
\begin{itemize}
	\item \textbf{Utilisateurs}
	
	\vspace{0.3cm}
	La participation complète dans les réseaux sociaux en ligne exige, que les utilisateurs enregistrent une identité (pseudo), bien que certains sites permettent la consultation de données publiques sans connexion explicite.
	
	\vspace{0.3cm}
	Les utilisateurs peuvent proposer des informations sur eux-mêmes (par exemple, leur date de naissance, leur lieu de résidence, leurs intérêts,...etc.) ce qui constitue le \textit{profil} de l'utilisateur.
	
	\vspace{0.3cm}	
	Le réseau social lui-même est composé de liens entre utilisateurs. Certains sites permettent aux utilisateurs à lier aux autres (sans le consentement du lien destinataire), tandis que d'autres sites suivent une procédure en deux phases, qui permet seulement de créer un lien lorsque les deux parties sont d'accord. 
	
	\vspace{0.3cm}	
	Les utilisateurs lient à d'autres  pour de nombreuses raisons. La cible d'un lien peut être une connaissance  du monde réel, un contact d'affaires, une connaissance virtuelle, quelqu'un, qui partage les mêmes intérêts, autre, qui télécharge un contenu intéressant, et ainsi de suite.
	
	\newpage	
	\item \textbf{Groupes}
	
	\vspace{0.3cm}
	La plupart des sites permettent également aux utilisateurs de créer des \textit{groupes} d'intérêts spéciaux. 
	
	\vspace{0.3cm}
	Les utilisateurs peuvent envoyer des messages à des groupes (visibles par tous les membres du groupe) et même télécharger du contenu partagé dans le groupe. Certains groupes sont supervisés et l'admission dans le groupe est contrôlée par un responsable du groupe unique, tandis que d'autres groupes sont ouverts à tous les membres.
	
	\vspace{0.3cm}	
	Tous les sites nécessitent aujourd’hui, une déclaration explicite des groupes par les utilisateurs, dont ils sont membres; ils doivent créer manuellement les groupes, nommer des administrateurs (si nécessaire).
	
	\vspace{0.3cm}
	La principale utilisation des groupes dans les réseaux actuels, consiste à exprimer des politiques de contrôle d'accès ou à fournir un forum pour le contenu partagé.
	
	\vspace{0.3cm}
	\item \textbf{Contenu}
	
	\vspace{0.3cm}
	Une fois l'identité créée, les utilisateurs de sites de partage de contenu peuvent télécharger du contenu sur leurs comptes. Beaucoup de ces sites, permettent aux utilisateurs de marquer le contenu, comme étant public (visible par tous) ou privé (visible uniquement par leurs «\textit{amis}»), et de marquer le contenu avec étiquettes.
	
	\vspace{0.3cm}
	De nombreux sites, tels que \textit{YouTube}, permettent aux utilisateurs de télécharger une quantité illimitée de contenu, tandis que d'autres sites, tels que \textit{Flickr}, exigent que les utilisateurs paient des frais d'abonnement ou soient soumis à une limite de téléchargement. 
	
	\vspace{0.3cm}
	Tout le contenu téléchargé par un utilisateur donné est répertorié dans son \textit{profil}, permettant aux autres de parcourir le réseau social pour découvrir de nouveaux contenus. En général, le contenu est indexé automatiquement et, s'il est accessible au public, rendu accessible via une recherche textuelle.
		
\end{itemize}

\section{Facebook vs. Twitter}

\begin{itemize}
	\item 
	Les réseaux Facebook, interconnectent les gens; tandis que les réseaux Twitter, interconnectent les idées et les sujets.
	
	\item Facebook, vous permet d'écrire un livre (sans que personne ne le lise). Twitter, est limité à 140 caractères par tweet.
	
	\item Facebook permettant plus d'options est considéré plus difficile à utiliser que Twitter.
	
	\item Facebook permet des \textit{Likes} et des \textit{Friends} tandis que l'appel à l'action de Twitter est \textit{Follow}.
	
	\item À l'intérieur de Facebook, vous \textit{aimerez} (\textit{Like}) ou \textit{partagerez} (\textit{Share}) quelque chose. Pourtant, à l'intérieur de Twitter, vous allez \textit{ReTweet} ou \textit{Favorite} quelque chose. 
	
	\item Facebook et Twitter permettent l'utilisation de \textit{Hashtags} pour regrouper des idées et des sujets. Facebook a incorporé l' idée de \textit{Hashtags} à l'image de Twitter.
	
	\item Vous pouvez rechercher dans l'un ou l'autre réseau pour des sujets, des personnes, des entreprises et des organisations.
	
	\item Les deux réseaux sont capables de permettre une certaine personnalisation pour inclure votre image de marque.

\end{itemize}

\vspace{-0.5cm}
\section{Le problème de diffusion des rumeurs dans les réseaux sociaux}
\vspace{-0.3cm}
 Un des problèmes les plus recherchés dans ces dernières années est le problème de la \textbf{détection des rumeurs et l'étude de sa propagation dans les réseaux sociaux.}
 
\vspace{0.2cm}	
Cependant, alors que les réseaux sociaux donnent l'accès à une source d'information sans précédent, l'absence d'efforts systématiques de la part des plateformes pour superviser les publications, conduit à la propagation de désinformation ou des rumeurs, qui nécessite un effort supplémentaire pour établir leur provenance, leur véracité et de quantifier leur taux de propagation.

\section{Définition et caractérisation des rumeurs}

\subsection{Définition des rumeurs dans la littérature}

Des publications récentes dans la littérature de recherche ont utilisé des définitions de rumeurs qui diffèrent les unes des autres. Par exemple :

\vspace{0.3cm}
Certains travaux récents ont défini une rumeur comme \textit{ " Une information jugée fausse " } \cite{Caietal.2014,Liangetal.2015}.

\vspace{0.3cm}
Tandis que la majorité de la littérature définit les rumeurs comme \textit{" Déclarations d'informations non vérifiées et pertinentes d'un point de vue instrumental en circulation"}\cite{DiFonzo.2007}.

\vspace{0.3cm}
Les principaux dictionnaires définissent une rumeur comme
,\textit{" Une histoire en cours ou un rapport de vérité incertaine ou douteuse "}  le dictionnaire anglais d'oxford\footnote{https://en.oxforddictionaries.com/definition/rumour}.

\vspace{0.3cm}
Une rumeur peut se terminer de trois façons: elle peut être résolue comme \textit{vraie} (crédible), \textit{fausse} (non crédible) ou \textit{non résolue}. 

\vspace{0.3cm}
Par conséquent, nous adhérons à cette définition dominante de la rumeur, qui la classe comme \textit{«Un élément d'information qui n'a pas encore été vérifié, et par conséquent sa valeur de vérité reste non résolue pendant qu'elle circule».}

\vspace{-0.4cm}
\subsection{Définition des rumeurs dans les réseaux sociaux}

Pour mieux enquêter sur la rumeur dans les sites de réseaux sociaux, nous la classifions en fonction de l'intention de l'utilisateur qui la diffuse :
\vspace{0.3cm}
\begin{itemize}
	\item[] \textbf{Dissémination involontaire de la rumeur :}
	
	Certaines informations erronées sont créées et transmises spontanément. Les gens ont tendance à contribuer à diffuser de telles informations, en raison de confiance en leurs amis et influenceurs dans un réseau social, et veulent informer leurs amis de ces informations.
	
	\item[] \textbf{Dissémination intentionnelle de la rumeur :}
	
	Certaines rumeurs et faux news, sont crées et diffusées intentionnellement par des utilisateurs mal intentionnés, pour susciter les préoccupations du public, tromper les gens et les utilisateurs des réseaux sociaux pour des profits abusifs.
	
\end{itemize}

\section{Définition de la problématique}

La problématique est de trouver un moyen qui calcule le taux de propagation des rumeurs sur les réseaux sociaux, afin de prédire la profondeur de la diffusion et pouvoir intervenir et les surveiller aux futures. 

\vspace{0.3cm}
Le but de notre contribution dans ce projet, est de quantifier, par simulation, le taux de propagation d’une rumeur sur les réseaux sociaux.

\vspace{-0.3cm}
\section{Conclusion}

Les premiers réseaux sociaux ont été créés au début des années 1990, mais il a fallu plusieurs années pour que ce nouveau type de communication sociale devienne une partie établie de la société Internet moderne.

\vspace{0.3cm}
Ce qui a suivi, a été un grand pas en avant dans l'histoire d'Internet, les personnes ayant des antécédents, des religions et des intérêts différents, ont décidé de se joindre à ces nouveaux médias, qui permet aux utilisateurs de choisir parmi un large éventail d'options différentes, afin de se décrire aussi précisément que possible.

\vspace{0.3cm}
Cependant, avec une liberté croissante d'utilisation et de diversité, vient le fardeau d'avoir à gérer de grande masse de données partagées, qui conduit à la propagation des rumeurs.

\vspace{0.3cm}
Nous avons aussi présenté dans ce chapitre la notion de rumeur et sa propagation dans les réseaux sociaux. Cette propagation et sa profondeur sont justement le sujet de recherche de notre contribution dans ce projet.

%% file: Chapitres/Chapitre-02.tex
\chapter{État de l'art}

\section{Introduction}

L'augmentation rapide des services de réseaux sociaux au cours des dernières années a permis aux gens de partager et de rechercher des informations de manière efficace. Cependant, Ces réseaux sociaux sont l'un des canaux les plus efficaces pour la désinformation. Compte tenu de la vitesse de diffusion de l'information dans ces derniers. Couplées à la propagation généralisée de fake-news, les rumeurs s'intensifient et peuvent avoir un impact significatif sur les utilisateurs avec des conséquences indésirables et faire des ravages instantanément.

\vspace{0.3cm}
Dans ce chapitre, nous citons les différents modèles et méthodes proposés pour poursuivre la trace de la propagation des rumeurs dans les réseaux sociaux.

\section{Modélisation de la rumeur} 

La rumeur peut être largement diffusée en très peu de temps. Pour expliquer comment l'information est diffusée dans un environnement en réseau, nous introduisons d'abord plusieurs modèles de diffusion.

\subsection{Diffusion de l'information dans les réseaux sociaux}

La diffusion de l'information dans les réseaux sociaux, peut être considérée comme un processus par lequel les news, les événements et les différents types d'informations sont affichés, transmis et reçus par les utilisateurs des réseaux sociaux. En représentant chaque utilisateur comme un nœud et l'amitié entre les utilisateurs comme des liens. Les réseaux sociaux sont transformés en un graphe $ G = (V, E) $, où $ V $ est l'ensemble des nœuds et $  E $ est l'ensemble des liens entre les nœuds. Ensuite, le processus de diffusion d'information peut être vu comme un signal ou une étiquette qui se propage dans le réseau.

\vspace{0.3cm}
Il existe différents modèles conçus pour abstraire le modèle de diffusion de l'information. Nous présentons ici trois modèles de diffusion, à savoir le modèle \textbf{SIR} (susceptible, infecté, récupéré) \cite{Kermack700}, le modèle de \textbf{basculement} \cite{Centola1194} et le modèle \textbf{en cascade indépendant} \cite{Kempe}.

\vspace{0.3cm}
Comme discuté dans \cite{Zafaran}, il y a trois rôles essentiels pour la diffusion:
\begin{enumerate}
	\item Les \textbf{expéditeurs} qui lancent le processus de la diffusion.
	\item Les \textbf{diffuseurs} qui transmettent ces informations à leurs abonnés.
	\item Les \textbf{destinataires} qui reçoivent des informations diffusées sur les sites des réseaux sociaux.
	
\end{enumerate}

Les principaux points de différenciation des différents modèles de diffusion sont :

 \begin{itemize}
 	\item La méthode d'information est diffusée entre les expéditeurs, les diffuseurs et les récepteurs. 
 	
 	\item L'évolution des rôles individuels au cours du processus de diffusion.
 	
 \end{itemize}

\subsubsection{Le modèle SIR}

Le modèle \textit{SIR}, décrit le processus de diffusion de l'information dans le réseau comme une propagation de maladie infectieuse dans une communauté. Ainsi, les nœuds sont généralement classés en trois catégories: 
\begin{enumerate}
	\item $ S $  \textit{susceptibles} d'être infectés;
	\item $ I $ les individus \textit{infectés}, qui sont actifs pour infecter d'autres;
	\item $ R $ les individus \textit{récupérés}, qui se sont rétablis et sont vaccinés contre la maladie. 
\end{enumerate}

\vspace{0.3cm}
Dans le contexte de la diffusion de l'information dans un réseau social, les nœuds \textit{infectés} peuvent être considérés comme ceux qui ont déjà été informés de certaines nouvelles ou d'événements, et sont prêts à les transmettre à leurs voisins; les nœuds \textit{récupérés} sont ceux qui ont été informés, mais ne transmettent pas l'information aux voisins; Les nœuds \textit{susceptibles} sont les utilisateurs qui ne sont pas informés, mais qui peuvent être informés par d'autres.

\vspace{0.3cm}
Ce modèle suppose, qu’un individu dans l’état $ R $ conserve son immunité. Par conséquent, un individu récupéré ne peut pas être de nouveau susceptible ou infecté.

\vspace{0.3cm}
Selon la catégorisation de l'utilisateur, l'échange d'informations se produit entre les nœuds \textit{infectieux} et les nœuds \textit{susceptibles}. Afin de modéliser le processus, un paramètre global est introduit comme la probabilité qu'un utilisateur susceptible soit activé; s'il a un ami infecté appelé $\beta$, en plus, un paramètre global est également introduit pour représenter la probabilité que des nœuds infectés soient récupérés $\Gamma $. La figure 2.1 illustre la structure du modèle \textit{SIR}. 

\vspace{0.3cm}
\begin{center}
	\includegraphics[scale=0.35]{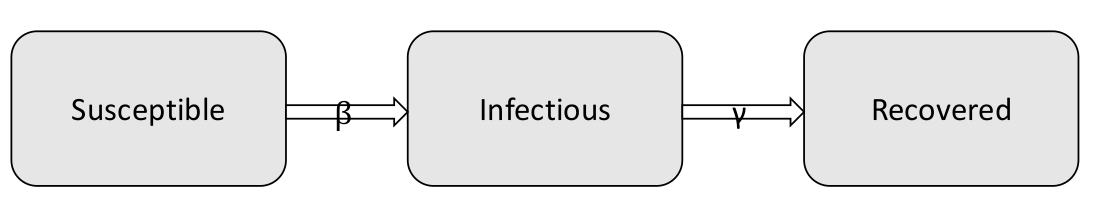}
	\vspace{-0.3cm}
	\captionof{figure}{Différents rôles d'utilisateur et leurs relations dans le modèle SIR.}
\end{center}

Dans le modèle SIR, si un utilisateur a plusieurs voisins infectieux liés, les liens sont tous indépendants les uns des autres et l'utilisateur peut être infecté au plus par un voisin. 
\newpage
Le processus d'infection est modélisé comme l'équation 1.1:
\vspace{-0.3cm}
\begin{equation}
I _{\beta}(a)=  1(\sum\limits_{(b,a)\in E,\\b\in V\cup I} 1 (f^{rand} \geq \beta) \textgreater 0 )
\end{equation}

Où $I_{\beta}(a)$ représente l'état d'infection d'un nœud sensible $ a $ dans l'horodatage suivant donné $ \beta $, et $ 1 (·) $ est une fonction qui est égale à \textit{un}, lorsque sa composante est vraie et nulle sinon. $ E $ et $ V $ sont l'ensemble des arêtes et des nœuds, respectivement, et $ I $ est l'ensemble des nœuds infectieux. Nous utilisons ($ b $, $ a $) pour désigner le lien \textit{directe / indirecte} entre deux nœuds et en utilisant \textbf{$ f^{rand} $} pour désigner la fonction de génération de probabilité aléatoire. Ainsi, le nœud $ b $ est le voisin infectieux de $ a $ et $ a $ sera activé si l'un de ses voisins l'active.

\subsubsection{Le modèle de basculement (Tipping)}

Comme le laisse entendre «The Power of Context», le comportement humain est fortement influencé par son environnement. Une intuition similaire a été adoptée par le modèle du point de basculement. Dans le modèle de basculement, il y a deux types d'utilisateurs: 
les personnes qui ont adopté le comportement et les personnes qui ne l'ont pas fait, où les diffuseurs et les expéditeurs ne sont pas explicitement identifiés. 

\vspace{0.3cm}
Puisque le processus d'adoption est irréversible, la diffusion de l'information ne se produit qu'entre la deuxième classe d'utilisateurs et leurs voisins actifs. Comme le montre la figure 2.2, un de ses amis sera influencé par l'adoption d'un certain comportement.

\begin{center}
	\includegraphics[scale=0.30]{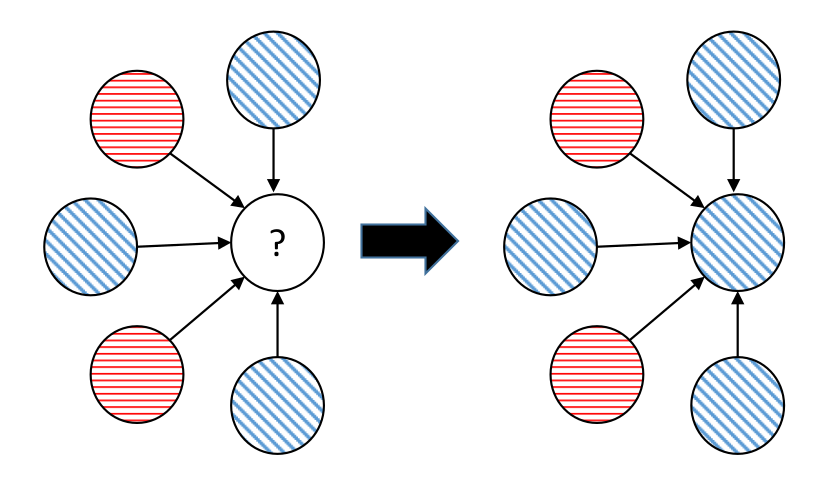}
	\vspace{-0.3cm}
	\captionof{figure}{un exemple illustratif du modèle Tipping.}
\end{center}

Le comportement peut être l'achat d'un nouveau téléphone portable ou porter une marque spécifique de vêtements. Afin de modéliser le processus de diffusion, une probabilité seuil $\theta$ est introduite pour juger, si le comportement atteint le «point de basculement»; et si le rapport des amis d'un utilisateur adoptant un certain comportement atteint la probabilité de basculement $\theta$, l'utilisateur adoptera également le comportement . 

\vspace{0.3cm}
Le processus d'infection est modélisé comme l'équation 1.2:

\begin{equation}
I _{\theta}(a)=  1(\sum\limits_{(b,a)\in E,\\b\in V\cup I} f(a,b) \geq \theta)
\end{equation}

Où $f(b, a)$ est la probabilité d'activation entre $ b $ et $ a $, et $ a $ sera activé lorsque l'influence de tous les voisins infectieux dépasse le seuil $\theta$.

\subsubsection{Le modèle en cascade indépendant}

Le modèle \textit{SIR} et le modèle de \textit{basculement} ont des différentes définitions de l'information diffusée, mais ils supposent tous qu'un paramètre global devrait fonctionner pour l'ensemble du réseau. Cette hypothèse forte réduit la complexité de calcul, mais ne parvient pas à gérer les situations complexes dans la réalité. 

Des modèles plus généralisés avec des paramètres variables sont proposés pour gérer de tels cas. Le modèle en\textbf{ Cascade Indépendant (CI)} est la forme généralisée du modèle \textit{SIR}. 

De même, il formule le processus de diffusion en tant qu'épidémie de maladie, mais la probabilité infectieuse est associée à différents liens. 

\vspace{0.3cm}
Le processus d'infection est modélisé comme l'équation 2.3:

\begin{equation}
I _{\beta}(a)=  1(\sum\limits_{(b,a)\in E,\\b\in V\cup I} 1 (f(a,b) \geq \beta) \textgreater 0 )
\end{equation}

La probabilité $ f(b, a) $ peut être obtenue en fonction de différentes applications; telles que la fréquence d'interaction, la proximité géographique,...etc. Ainsi, le modèle \textbf{CI} est capable d'intégrer plus d'informations contextuelles.

\subsection{Diffusion de la rumeur}

La diffusion de la rumeur est davantage liée à la confiance et à la croyance dans les réseaux sociaux. Les modèles épidémiques, y compris \textit{SIR} et \textit{CI}, supposent que l'infection se produit entre un utilisateur infectieux et un utilisateur susceptible avec une probabilité prédéfinie. Comme mentionné précédemment, la probabilité peut augmenter avec plus d'interactions ou d'autres conditions contextuelles. Bien que les liens d'un utilisateur soient indépendants, un utilisateur qui a plus d'amis infectieux est plus susceptible d'être infecté. Le modèle \textit{Tipping} contient également un paramètre permettant d'estimer la probabilité d'infection d'un utilisateur en fonction du nombre d'amis activés.

\vspace{0.3cm}
De tels phénomènes révèlent que, quel que soit le degré de diffusion d'une information erronée, certains nœuds ne seront pas affectés et continueront à intervenir dans une telle diffusion. Afin de modéliser un tel processus, l'approche existante catégorise les gens en deux genres: les individus réguliers et les individus énergiques.

\vspace{0.3cm}
Le processus de diffusion de l'information est reformulé en échange de croyance. Un paramètre $ \theta \in R $ est introduit pour représenter la probabilité d'apprendre ou d'accepter certaines informations. Le processus de diffusion de la rumeur est alors défini comme l'échange de croyance entre deux nœuds.

\begin{center}
	\includegraphics[scale=0.30]{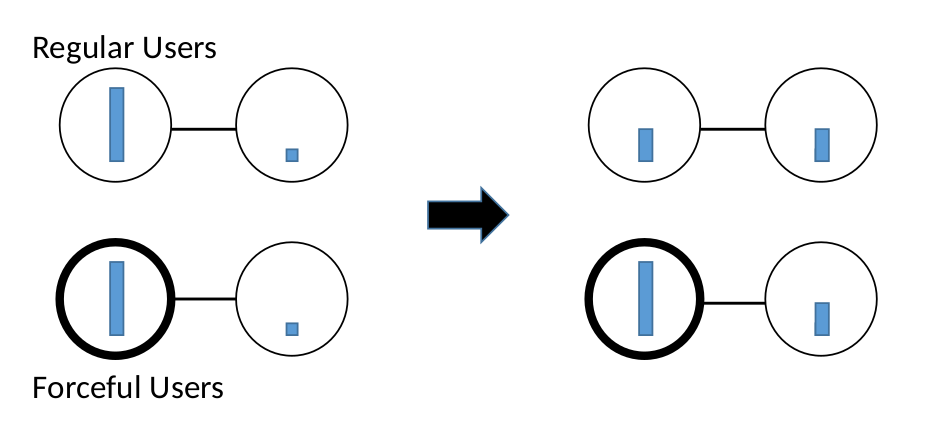}
	\captionof{figure}{Un exemple d'échange de croyance de rumeur, où la hauteur des barres dans les cercles représente l'étendue de la croyance.}
\end{center}

\newpage
Comme l'illustre la figure 2.3, les échanges entre différents utilisateurs sont différents. Lorsqu'un utilisateur régulier interagit avec un autre utilisateur régulier, ils se touchent simultanément. Une croyance consensuelle sera obtenue en faisant la moyenne de leurs croyances. Mais quand un nœud régulier interagit avec son homologue puissant, seul le nœud régulier sera affecté et l'individu énergique gardera la croyance originale.

\vspace{0.3cm}
Au début de la diffusion de la rumeur, tous les nœuds réguliers sont supposés avoir une probabilité de croyance tirée d'une certaine distribution de probabilité, et les individus puissants peuvent détenir des scores spécifiques par rapport à la normale. La croyance converge vers un consensus parmi tous les individus, par des itérations d'échange de croyances. Plus la croyance consensuelle est élevée, plus la rumeur répandue est diffusée sur le réseau. Plus formellement, le processus est modélisé comme suit:

\begin{equation}
	x_{i}(0)=  \epsilon_{i} x_{i}(0) + (1- \epsilon_{i} x_{j}(0))
\end{equation}

Ici, nous supposons que le réseau social a $ n $ agents. $x(0)\in R^{n}$ enregistre le niveau de croyance des utilisateurs à l'instant 0. L'équation 2.4 montre comment l'échange de croyance a lieu entre deux nœuds $ i $ et $ j $, où $ \epsilon $ représente la mesure dans laquelle l'utilisateur $ i $ reste à sa croyance originale. 

\vspace{0.3cm}
Si l'utilisateur $ i $ et $ j $ sont des utilisateurs réguliers, alors $\epsilon_{ i } = 0.5 $ et un tel changement se produit au score de croyance de $ j $ ($ \epsilon_{ i } = 0.5 $). Si seul l'utilisateur $ i $ régulier mais l'utilisateur $ j $ est un individu puissant, alors $ \epsilon  \textless 0.5 $ et l'utilisateur $ j $ n'est pas significativement affecté par l'échange ($ \epsilon_{ j } \approx 1 $). Lorsque l'utilisateur $ i $ est un utilisateur puissant, alors $ \epsilon \approx 1 $ et l'utilisateur $ j $ ne seront affectés que s'il est régulier.

\section{Les différentes applications sur la propagation des rumeurs}

\cite{doi,Shah} proposent des travaux qui se concentrent sur la propagation de rumeurs à travers le réseau social. Ils essaient d'utiliser la théorie des graphes pour détecter et trouver la source des rumeurs.

\vspace{0.3cm}
D'autres types de systèmes sont plus un système de surveillance pour suivre la distribution de maladies infectieuses , utilisant des modèles \textbf{épidémiologiques} à partir des signaux linguistiques sur le Web \cite{doi44}.

\vspace{0.3cm}
\cite{Spiro} ont également modélisé le taux de postes au fil du temps dans leur exploration de la rumeur, lors de la marée noire de Deepwater Horizon en 2011.

\vspace{0.3cm}
\cite{Kwono} ont promu des utilisations des caractéristiques temporelles, structurelles et linguistiques. Les caractéristiques linguistiques sont liées au plus grand nombre de mots utilisés dans les messages et sont tirées d'un dictionnaire de sentiment (4500 mots). Les caractéristiques temporelles soulignent la périodicité du phénomène de la rumeur et donnent de l'importance à un choc externe qui peut entraîner non pas un, mais plusieurs impacts au fil du temps.

\vspace{0.3cm}
Le site \textbf{TwitterTrails}\footnote{http://twittertrails.com} est l'un des rares outils qui ne présente pas seulement une base de données, mais aussi une exploration intelligente des informations (chronologie, propagateurs, négation, rafale, initiateur, acteurs principaux) dans 547 articles de réseaux sociaux.

\vspace{0.3cm}
Le projet «truthy» de l'Université d'Indiana des États-Unis, consiste a crée une plate-forme «OSoMe» (Observatory on Social Media) étudier la diffusion de l'information en ligne et discriminer entre les mécanismes qui favorisent la propagation des rumeurs sur les réseaux sociaux.
\newpage
Dans ce dernier projet, ils collectent des données volumineuses à partir de flux de microblogs publics et analysent le partage d'informations à l'aide d'outils et de modèles de réseaux complexes.

\vspace{0.3cm}
\cite{Budak} prouvent que minimiser la propagation de la rumeur dans les réseaux sociaux est un problème NP-difficile et fournit également une solution approchée gourmande.

\vspace{-0.5cm}
\section{Critiques}

Différents modèles ont été proposés pour étudier les caractéristiques comportementales des graphes mais pas vraiment des personnes. La propagation des rumeurs a également été considérée comme un modèle dynamique, où les acteurs sociaux sont classés comme sensibles, infectés et récupérés; mais ne jamais citer en quoi se base cette classification.

\vspace{0.3cm}
Nous pouvons voir le contenu d'une rumeur comme un document. De cette façon, la formulation du problème est un peu différente car elle concerne la description d'un contenu et sa relation avec d'autres contenus évoluant dans le temps qui représentent l'historique de partage de chaque utilisateur. Donc, la propagation de cette rumeur peut être étudié en basant sur les caractéristiques comportementales des personnes et leurs connexions. 

\vspace{0.3cm}
En effet, historiquement c'est exactement ce que les sociologues ont fait, en étudiant les individus et les groupes sociaux. 
\vspace{-0.4cm}  
\section{Conclusion}

La diffusion des rumeurs sur les réseaux sociaux est un sujet très récent. Dans ce chapitre, nous avons présenté les différents modèles de diffusion des rumeurs sur les réseaux sociaux, ensuite nous avons cité quelques applications qui sont déjà faites dans ce domaine.

\vspace{0.3cm}
Il faut rappeler dans ce contexte que, vu que le sujet est très récent,  il existe peux de travaux académiques ayant un lien direct avec notre sujet proposé.
 
\vspace{0.3cm}
Dans le prochain chapitre, nous proposerons notre propre vision dans le domaine cité. Il s'agira d'une nouvelle méthode de quantification de la propagation des rumeurs.

%% file: Chapitres/Chapitre-03.tex
\chapter{Une approche pour la quantification de la profondeur d'une rumeur}

\section{Introduction}
L'émergence des réseaux sociaux au cours de la dernière décennie, a conduit à une augmentation considérable du volume d'informations sur les individus, leurs activités, les liens entre ces derniers et les groupes auxquels ils appartiennent , leurs opinions et ainsi que leurs pensées. Une grande partie de ces données peut être modélisée sous forme d'étiquettes (caractéristiques) associées à des individus; pouvant générer leurs profils, qui sont à leur tour représentés sous la forme de nœuds dans un graphe. Ces caractéristiques peuvent prendre plusieurs formes: 
\vspace{0.3cm}
\begin{itemize}
	\item Des caractéristiques démographiques, telles que l'âge, le sexe et l'emplacement;
	\item Des caractéristiques qui représentent des croyances politiques ou religieuses;
	\item Des caractéristiques qui codent les intérêts, les hobbies et les affiliations;
	\item Beaucoup d'autres caractéristiques possibles, capturant des aspects des préférences ou du comportement d'un individu. 
\end{itemize}

\newpage
Les caractéristiques apparaissent généralement sur le profil de l'utilisateur dans le réseau ou sur d'autres objets du réseau (photos, vidéos,...etc.).

\vspace{0.3cm}
Dans ce chapitre, nous proposons notre contribution qui consiste en la quantification de la propagation des rumeurs sur les réseaux sociaux et ce, en se basant sur l'utilisation des profils des acteurs sur ces réseaux en capturant leurs aspects des préférences ou du comportement de chacun de ces acteurs. 

\section{Notre contribution}

Dans cette contribution, nous nous attaquons au problème de la prédiction de la diffusion des rumeurs dans les réseaux sociaux. Nous avons proposé une nouvelle étude basée sur la notion de similarité; afin de calculer la probabilité de diffuser un contenu donné, en étudiant l'effet de la similarité :
\begin{itemize}
	\item Utilisateur-utilisateur.
	\item Utilisateur-contenu.
\end{itemize}
 sur la propagation de la rumeur. Et ce, en prenant en compte :
\begin{itemize}
	\item Le contenu de l’information diffusée.
	\item Les goûts des utilisateurs et leur intérêt (sujets).
\end{itemize}
 
\subsection{Calcul de similarité}
 
 Une mesure de similarité est, en général, une fonction qui quantifie le rapport entre deux objets, comparés en fonction de leurs points de ressemblance et de dissemblance. Les deux objets comparés sont, bien entendu, de même type.
 
 \vspace{0.3cm} 
 Notre approche, dans cette contribution, est basée sur la similarité textuelle, entre les sujets qui concernent l'historique de partage de chaque utilisateur; donc nous avons jugé utile d'étudier le problème en testant les mesures de similarité les plus efficaces. Nous pouvons en citer les suivantes :
 
\newpage
 \begin{enumerate}
 	\item \textbf{La similarité Cosinus :}
 	
 	 \vspace{0.3cm} 
 	La similarité cosinus, est fréquemment utilisée en tant que mesure de ressemblance entre deux documents $d_{1}$ et $d_{2}$ . Il s'agit de calculer le cosinus de l'angle entre les représentations vectorielles des documents à comparer \cite{Baeza-Yates}.
 
 	\begin{equation}
 	sim_{cosinus} (d_{1} , d_{2} ) = \frac{\vec{d_{1}}\cdot\vec{d_{2}}}{ \lVert d_{1} \rVert\lVert d_{2} \rVert} \in [0, 1]	
 	\end{equation} 

 	\item \textbf{La similarité de corrélation de Pearson :}
 	
 	 \vspace{0.3cm} 
 	La similarité de corrélation de Pearson, calcule la similarité entre deux documents $d_{1}$ et $d_{2}$ comme le cosinus de l'angle entre leurs représentations vectorielles centrées-réduites.
 	 
 	\begin{equation}
 	sim_{pearson} (d_{1} , d_{2} ) = sim_{cosinus} (d_{1} - \bar{d_{1}} , d_{2} - \bar{d_{2}} ) \in [-1, 1]	
 	\end{equation} 
 	Où $ \bar{d_{1}} $ (resp. $ \bar{d_{2}} $ ) représente la moyenne de $ d_{1} $ (resp. $ d_{2} $ ).
 	
 	\item \textbf{La similarité de Jaccard :}
 	
 	 \vspace{0.3cm} 
 	L'indice de Jaccard ou coefficient de Jaccard, est le rapport entre la cardinalité (la taille) de l'intersection des ensembles considérés et la cardinalité de l'union des ensembles. Il permet d'évaluer la similarité entre les ensembles. Les documents $d_{1}$ et $d_{2}$ sont donc représentés, non pas comme des vecteurs, mais comme des ensembles de termes \cite{jaccard1901etude}. 
 	  
 	\begin{equation}
 	sim_{jaccard} (d_{1} , d_{2} ) = \frac{\lVert d_{1} \cap d_{2}\rVert}{ \lVert d_{1} \cup d_{2}\rVert} \in [0, 1]	
 	\end{equation} 
 	Il est aussi possible d'utiliser la représentation vectorielle.
 	
 	\begin{equation}
 	sim_{jaccard} (d_{1} , d_{2} ) = \frac{\vec{d_{1}}\cdot\vec{d_{2}}}{ \lVert d_{1} \rVert\lVert d_{2} \rVert - \vec{d_{1}}\cdot\vec{d_{2}} } \in [0, 1]
 	\end{equation}
 	
 	 \newpage
 	\item \textbf{La similarité de Levenshtein (d'édition) :}
 	
 	 \vspace{0.3cm} 
 	La similarité de Levenshtein, calcule la similarité entre les représentations sous forme de chaines de caractères des documents $ d_{1} $ et $ d_{2} $ . Il s'agit du coût minimal; c'est-à-dire du nombre minimal d'opérations d'édition, pour transformer $ d_{1} $ en $ d_{2} $ \cite{Levenshtein_SPD66}. Les opérations sont les suivantes :

 	\begin{itemize}
 		\item La substitution d'un caractère de $ d_{1} $ en un caractère de $ d_{2} $.
 		\item L'ajout dans $ d_{1} $ d'un caractère de $ d_{2} $.
 		\item La suppression d'un caractère de $ d_{1} $.
 		
 	\end{itemize}
 		
 	 \vspace{0.3cm} 
 	Pour obtenir la similarité de Levenshtein $ sim levenshtein (d_{1} , d_{2} ) $ entre les documents  $ d_{1} $  et  $ d_{2} $ , il s'agit d'associer à chacune de ces opérations un coût. Le coût des opérations est toujours égal à 1, sauf dans le cas d'une substitution de caractères identiques. Notons que cette similarité a été étendue pour prendre en compte la grammaire, la phonétique,...etc.

 	\item \textbf{La similarité Dice :} 
 	
 	 \vspace{0.3cm}  	
 	L'indice de Dice, mesure la similarité entre deux documents $d_{1}$ et $d_{2}$ en se basant sur le nombre de termes communs à $d_{1}$ et $d_{2}$.
 	  
 	\begin{equation}
 	sim_{dice} (d_{1} , d_{2} ) = \frac{2 N_{c}}{N_{1} + N_{2}} \in [0, 1]
 	\end{equation}
 	 \vspace{0.3cm} 
 	Où $ N_{c} $ est le nombre de termes communs à $ d_{1} $ et $ d_{2} $ , et $ N_{1} $ (resp. $ N_{2} $ ) est le nombre de termes de $ d_{1} $ (resp. $ d_{2} $ ).

 \end{enumerate}
  	\vspace{-0.3cm}
 	\paragraph{\textbf{Comparatif}\\
 		}{
 	
 	\vspace{0.3cm}
 	Nous tentons maintenant d'identifier une étude comparatif selon les performances de chaque mesure.  \cite{Huang08} et \cite{Strehl} ont tous les deux montré que les performances de la similarité cosinus, du coefficient de Jaccard et de Pearson sont très proches. Cependant, \cite{bavi} fait apparaitre que plus le document est de petite taille, mauvais sont les résultats avec la similarité cosinus ou avec le coefficient de Jaccard. 
 	
 	\vspace{0.3cm}
 	Sachant que l'indice de Dice est fonction du Jaccard, nous pouvons penser qu'ils ont des performances similaires. La distance de Levenshtein est largement utilisée en linguistique et en bio-informatique ainsi que pour la reconnaissance de blocs de textes contenants des erreurs isolées. 
 	
 	\vspace{0.3cm}
 	Malheureusement, le temps de calcul  (complexité), lorsqu'on l'applique à deux séquences d'approximativement la même taille, $ n $, est $ O(n^{2}) $. Cela est un obstacle dans de nombreuses applications pratiques \cite{baake}.
 	
	}
 
 \subsection{Algorithmes de diffusion de notre contribution}
 
D’après la proposition de notre étude de la quantification de la propagation des rumeurs dans les réseaux sociaux cités ci-dessus, nous proposons nos algorithmes comme suit :

\vspace{0.3cm}
Un réseau social est modélisé sous la forme d'un graphe $G = (V, E)$ où $ V $ est l'ensemble de tous les utilisateurs  et $ E $ est l’ensemble d' arrêtes où chaque arrête représente la relation entre deux individus.

\vspace{0.3cm} 
L'algorithme suivant présente la relation utilisateur-utilisateur :

 \vspace{0.3cm} 
 \begin{algorithm}[!ht]
 	\caption{  Algorithme de diffusion des rumeurs ( Utilisateur-Utilisateur).}
 	\begin{algorithmic} 
 		\REQUIRE  
 		Un graphe $ G = (V, E) $ ; Un ensemble d’utilisateurs diffuseurs \textit{Initiaux}.
 		
 		\ENSURE Le nombre des utilisateurs \textit{Diffuseurs} après le processus de diffusion.
 		
 		\STATE 		\textit{Diffuseurs} $\leftarrow$  \textit{Initiaux};\\
 		
 		\FORALL{ $ e_{i} \in Diffuseurs $ }
 		\FORALL{$ e_{j} \in VoisinsSortantde(e_{i}) $  }
 		\IF{$ e_{j} \notin Diffuseurs $}
 		
 		\IF{ $ e_{i}$  \textit{ Similaire à}   $  e_{j}$}
 		
 		\STATE $ Diffuseurs \leftarrow  Diffuseurs $   $  \cup$ \{$  e_{j} $\};
 		
 		\ENDIF
 		\ENDIF
 		\ENDFOR
 		\ENDFOR
 		
 		\PRINT \textbf{taille}(\textit{Diffuseurs});
 		
 	\end{algorithmic}
 \end{algorithm}
\newpage

\textbf{Entrée :} Un graphe d’utilisateur, une liste des diffuseurs Initiaux.

\vspace{0.3cm} 
\textbf{Sortie :} Nombre d’utilisateur diffuseurs de la rumeur.

\vspace{0.3cm}
\textbf{Principe :}  Si $ U_{1} $ et $ U_{2} $ (deux utilisateurs voisins)  ont des similaire historique (en terme de sujets), et $ U_{1} $ est un diffuseur de la rumeur;  alors  $ U_{2} $ va être un diffuseur de la rumeur.

 \vspace{0.3cm}
Cet algorithme, une fois déployé, compte le nombre d'acteurs diffuseurs sur le réseau. Il est basé sur la notion de similarité entre les profils de ces acteurs.

 \vspace{0.3cm} 
Notons que le type de similarité qui sera choisi en fonction des résultats les plus  fiables dans  notre prochain cas d'étude. 

\vspace{0.3cm}
L'algorithme suivant présente la relation utilisateur-contenu\_rumeur :

 \vspace{0.3cm} 
 \begin{algorithm}[!ht]
 	\caption{ Algorithme de diffusion des rumeurs ( Utilisateur-Contenu).}
 	\begin{algorithmic} 
 		\REQUIRE  
 		Un graphe G = (V, E) ; Un ensemble d’utilisateurs diffuseurs \textit{Initiaux};
 		Un Contenue \textit{Rumeur} .
 		\ENSURE Le nombre des utilisateurs \textit{Diffuseurs} après le processus de diffusion.
 		
 		\STATE 		\textit{Diffuseurs} $\leftarrow$  \textit{Initiaux};\\
 		
 		\FORALL{ $ e_{i} \in Diffuseurs $ }
 		\FORALL{$ e_{j} \in VoisinsSortantde(e_{i}) $  }
 		\IF{$ e_{j} \notin Diffuseurs $}
 		
 		\IF{ $ e_{j}$  \textit{ Similaire à}   $ Rumeur $}
 		
 		\STATE $ Diffuseurs \leftarrow  Diffuseurs $   $  \cup$ \{$  e_{j} $\};
 		
 		\ENDIF
 		\ENDIF
 		\ENDFOR
 		\ENDFOR
 		
 		\PRINT \textbf{taille}(\textit{Diffuseurs});
 	\end{algorithmic}
 \end{algorithm}
 
 \textbf{Entrée :} Un graphe d’utilisateur, une liste des diffuseurs Initiaux.
 
 \vspace{0.3cm} 
 \textbf{Sortie :} Nombre d’utilisateur diffuseurs de la rumeur.
 
 \vspace{0.3cm}
 \textbf{Principe :} Si $ U_{1} $ et $ U_{2}$ (deux utilisateurs voisins), $ U_{1} $ est un diffuseur de la rumeur et $ U_{2}$ similaire au contenu de la rumeur (en terme de sujets);  alors $ U_{2}$ va être un diffuseur de la rumeur.
 
\newpage
 Dans ce dernier algorithme, nous quantifions la profondeur de la diffusion d'une rumeur entre acteurs de cette rumeur et ce, en se basant sur la similarité de contenu de la rumeur en l'associant à leurs profils.

 \section{Conclusion}
 
 Dans ce chapitre, nous avons proposé notre approche caractérisée par des algorithmes afin de  déterminer la profondeur de diffusion d'une rumeur.
 \vspace{0.3cm}

 Nous avons présenté aussi dans ce chapitre, les différents types de similarités les plus connus (Cosinus, corrélation de Pearson,  Jaccard, Levenshtein et Dice). Nous utiliserons une des métriques citées en fonction de la spécificité de notre cas d'étude.
 \vspace{0.3cm}
  
 Notre approche, sera déployée et validée dans le chapitre suivant, elle permettra de déterminer la fiabilité de notre logique et déterminer les résultats de notre proposition qui consistera en la quantification de la profondeur de diffusion des rumeurs.

%% file: Chapitres/Chapitre-04.tex
\chapter{Déploiement de l'approche proposée et résultats}

\section{Introduction}
Après avoir présenté notre approche académiquement sous la forme de deux algorithmes, utilisateur-utilisateur et utilisateur-contenu, afin de quantifier la profondeur d'une rumeur sur le réseau social, nous avons adopté le cas de Twitter comme terrain de déploiement de notre approche, après réflexion et comparaison avec d'autres réseaux existants.

\vspace{0.3cm}
Nous proposons dans ce chapitre, les différentes phases suivies lors de notre déploiement, afin de mettre en œuvre, valider notre approche et interpréter les résultats obtenus.  

\section{Étude de cas Twitter}

Soroush Vosoughi, Deb Roy et Sinan Aral ont étudié la manière dont des news, fausses et vraies, ont été diffusées sur Twitter entre 2006 et 2017. Ils ont analysé le parcours de \textit{126.000} d'entre elles, rediffusées plus de \textit{4,5 millions} de fois par 3 millions de personnes.

\newpage
Pour déterminer si les news étaient vraies ou fausses, les trois chercheurs ont fait appel à six organisations indépendantes spécialisées dans le fact-checking. Le résultat est ce que certains qualifient déjà comme "la plus grande étude longitudinale [suivie dans le temps] jamais réalisée sur la diffusion des fausses news en ligne ".

\vspace{0.3cm}
Selon cette étude publiée dans le magazine Science en mars 2018 \cite{Vosoughi1146}, les trois chercheurs trouvent que les fausses informations sont\textit{ \textbf{70\%}} plus susceptibles d'être partagées sur Twitter que les vraies et que les fausses informations ou "\textit{fake-news}" se propagent également plus largement que celles qui sont authentiques.

\vspace{0.3cm}
C'est sur cette base, que nous avons opté pour Twitter comme réseau social, afin de déployer notre approche.
\vspace{-0.2cm}
\subsection{Terminologie}
Ce présent travail, se concentre sur les rumeurs qui se propagent sur \textbf{Twitter}. En tant que tel, nous devons définir plusieurs termes spécifiques à \textbf{Twitter}; afin de présenter le vocabulaire et terminologie utilisés par ce réseau social.

\vspace{0.3cm}
\begin{itemize}
	\item \textbf{Retweet}
	
	Un retweet est un re-post ou un repartage d'un tweet par un autre utilisateur. Il est indiqué par les caractères RT.
	\item \textbf{Favorite}
	
	Les favoris sont utilisés par les utilisateurs lorsqu'ils aiment un tweet. En favorisant un tweet, un utilisateur peut laisser savoir à l'auteur qu'il a aimé son tweet. Le nombre total de fois qu'un tweet a été favorisé est visible pour tout le monde.
	\item \textbf{Verified User}
	
	Un utilisateur de Twitter \textit{"Verified User"} est un utilisateur que Twitter a confirmé être le vrai. La vérification est effectuée par Twitter pour établir l'authenticité des identités des individus et des marques clés. Le statut vérifié d'un utilisateur est visible pour tout le monde.
	\item \textbf{Followers }
	
	Les abonnés (Followers) d'un utilisateur sont d'autres personnes qui reçoivent les tweets et les mises à jour de l'utilisateur. Lorsqu'un utilisateur est suivi par quelqu'un, il apparaîtra dans sa liste d'abonnés. Le nombre total d'abonnés d'un utilisateur est visible par tous.
	\item \textbf{Followees}
	
	Les Followees d'un utilisateur sont d'autres personnes que l'utilisateur suit. Le nombre total de Followees d'un utilisateur est également visible par tous.
	\item \textbf{Follower Graph}
	
	Le graphe des utilisateurs sur Twitter et leurs relations suiveuse. Les nœuds du graphe sont des utilisateurs et les bords directionnels représentent la relation suiveuse entre les utilisateurs.
	
\end{itemize}

\section{Phases de déploiement de l'approche}

Afin d'obtenir des résultats fiables, nous avons utilisé des données réelles à partir du réseau social \textbf{Twitter}. 

\vspace{0.3cm}
Pour l'implémentation des algorithmes, nous les avons projetés en langage de programmation \textit{Python}; car ce dernier présente des avantages et des facilitations, qui nous arrangent lors du déploiement de la solution, ce choix nous a permis, en plus de nos propres codes de programmation de réutiliser un certain nombre d'API que nous allons présenter ultérieurement. 

\vspace{0.3cm}
Ces phases sont présentées comme suit:

\subsection{Phase de l'extraction des données}

Pour assurer que l'information voulue utiliser pour l'étude est une rumeur, nous avons choisie l'\textbf{API} de \textbf{Hoaxy}\footnote{http://hoaxy.iuni.iu.edu}.

Cette dernière \textbf{API}, permet de télécharger les \textit{Ids} (identificateurs) des utilisateurs \textbf{Twitter} qui diffusent cette rumeur.

\vspace{0.3cm}
À partir de cette plateforme, nous avons choisi une récente rumeur concernant la société de \textbf{Facebook} :\textit{" Mark Zuckerberg Prepares For Congressional Testimony By Poring Over Lawmakers’ Personal Data "}

\vspace{0.3cm}
Après le téléchargement des identificateurs des utilisateurs diffuseurs de la rumeur, nous avons utilisé l'\textbf{API} de \textbf{Twitter} ; qui offre un accès programmé pour lire et écrire des données sur le réseau social \textit{Twitter}. 
La façon la plus pratique de se familiariser avec cette \textbf{API}, est d'utiliser sa \textit{console} (site Web de la console API Twitter, 2015). Cette console permet à un navigateur Web d'envoyer des requêtes à l'\textbf{API Twitter} avec des méthodes. Un grand nombre de ces méthodes nécessitent que les utilisateurs choisissent une méthode d'authentification. Une fois que les \textit{URLs} correctes des demandes sont validées dans la console, l'utilisation d'un langage de programmation général pour programmer un script qui demande les données voulues est acceptée.

\vspace{0.3cm}
Nous avons choisi d'associer le langage de programmation \textbf{Python} avec ses requêtes et sa bibliothèque \textit{" Tweepy "}.
\vspace{0.3cm}
\begin{pyin}
import tweepy
# Initialisation des attributes d'authentification :
ACCESS_TOKEN = 'XXXXXX'
ACCESS_SECRET = 'XXXXXX'
CONSUMER_KEY = 'XXXXXX'
CONSUMER_SECRET = 'XXXXXX'
# Authentification : 
auth = tweepy.OAuthHandler(CONSUMER_KEY, CONSUMER_SECRET)
auth.set_access_token(ACCESS_TOKEN, ACCESS_SECRET)
# initialisation des attributes de l'API :
api = tweepy.API(auth,wait_on_rate_limit=True)
\end{pyin}
\vspace{-0.5cm}

\begin{center}
	\captionof{figure}{Code Python de processus d'authentification de l'\textbf{API Twitter} utilisant la bibliothéque \textbf{Tweepy}.}
\end{center}
\vspace{-1cm}

Les méthodes de l'\textbf{API} les plus pertinentes pour la collection (\textit{Scraping}) de données sur \textbf{Twitter} et leurs limites, indiquées ci-dessous:
\begin{itemize}
	\item \textit{statuses/retweet/{id}.json} :
	
	Cette opération obtient jusqu'à 100 retweets d'un identifiant de tweet donné. Par conséquent, il peut être utilisé pour étendre n'importe quel ensemble de données Twitter si les identifiants de tweets sont inclus. En outre, les données de l'utilisateur retweeteur sont également données dans la réponse. La principale limite est qu'elle ne récupère qu'un petit pourcentage du nombre total de retweets (jusqu'à 100). De plus, cette méthode ne permet que 15 requêtes toutes les 15 minutes. Par conséquent, obtenir des retweets prend beaucoup plus de temps que d'obtenir des tweets avec recherche. Un avantage d'obtenir des données Twitter à partir des tweets ou des identifiants d'utilisateurs avec cette ou les opérations suivantes est que ces données peuvent être récupérées après les 9 jours maximum que l'opération de recherche donnée.
	
	\item \textit{statuses/show/{id}.json} :
	
	Les conditions d'utilisation actuelles de Twitter stipulent, que si vous fournissez du contenu à des tiers, y compris des ensembles de données téléchargeables, vous ne diffuserez ou autoriserez le téléchargement que des IDs de Tweet. Une pratique assez courante dans le milieu de la recherche consiste à distribuer ouvertement des listes d'identifiants de tweet au lieu des données brutes. Cette opération permet aux chercheurs de récupérer le tweet et les données des utilisateurs à partir d'un identifiant. Comme avec l'opération \textit{retweet}, cette méthode ne permet que 15 requêtes toutes les 15 minutes.
	
	\item \textit{followers/ids.json} et \textit{friends/ids.json} :
	
	Ces opérations permettent aux chercheurs d'obtenir des abonnés et des amis à partir de l'identifiant d'un utilisateur (qui est dans les données de tweets données avec les opérations expliquées ci-dessus). Comme dans l'opération \textit{retweet}, la collecte de ces données est plus restrictive que la méthode de recherche: seulement 15 requêtes toutes les 15 minutes.
	
	\item \textit{lists/statuses.json} :
	
	Cette opération renvoie une chronologie des tweets créés par les utilisateurs spécifiés (historique de tweets), leurs retweets sont aussi inclus par défaut.

\end{itemize}

\vspace{0.3cm}

En utilisant \textbf{Python} avec le \textbf{package tweepy} connecté avec l'API, nous avons pu télécharger les tweets (historique des posts) de chaque utilisateur, en plus pour chaque utilisateur, nous avons téléchargé \textit{30} "suiveurs (Followers)" non-diffuseurs de la rumeur pour la validation des résultats.

\vspace{0.3cm}
Cette opération prend \textbf{15h 46min 7s} en utilisant une machine de \textit{6Gb} RAM et \textit{2Mb} comme vitesse d'Internet.

\subsection{Phase de traitement des données}
Dans cette étape, nous nous concentrons sur l'extraction des sujets à partir des historiques des tweets des utilisateurs. Où nous avons utilisé l'\textbf{API} de \textbf{MonkeyLearn}, qui prend comme \textit{input} un texte et comme \textit{output} ses différents sujets. 
\vspace{-0.3cm}
\paragraph{L'API MonkeyLearn : \\	}{

	C'est une plateforme d'\textit{IA} (Intelligence Artificielle) qui permet de classer et d'extraire des données exploitables à partir de textes bruts tels que des courriels, des chats, des pages Web, des documents, des tweets et plus encore! . Vous pouvez classer les textes avec des catégories ou des étiquettes personnalisées comme les sentiments ou les sujets, et extraire des données particulières comme des organisations ou des mots-clés.
}

\begin{pyin}
from monkeylearn import MonkeyLearn
ml = MonkeyLearn('XXXXX')
# Classificateur de detection des sujets :
module_id = 'cl_5icAVzKR'
r,c = users.shape
Topics = pd.DataFrame(index= np.arange(0,r),
columns=['user_topics'])
for i in range(0,r):
  # Sujets des Tweets :
  res = ml.classifiers.classify(module_id, [users.iloc[i,1]]
                                ,sandbox=False)
  topics =""
  for x in res.result[0] :
    topics = topics + " , " + x["label"]
  Topics.iloc[i] = topics    
	
\end{pyin}

\begin{pyout}
	time : 9h 33m  38s
\end{pyout}

\vspace{-0.7cm}

\begin{center}
	\captionof{figure}{Code Python d'extraction des sujets des tweets d'utilisateurs utilisant l'\textbf{API MonkeyLearn}.}
\end{center}
 \vspace{-1.5cm} 
 
\subsection{Phase de calcul de similarité}

Après les deux premières phases, nous avons obtenu comme résultats:
\vspace{0.2cm}
\begin{itemize}
	\item \textbf{\textit{17083}} utilisateurs avec ses sujets et dates de diffusion du rumeur.
	\item \textbf{\textit{25242}} liens entres utilisateurs.
	\item  et aussi les sujets du rumeur.
\end{itemize}
\vspace{0.1cm}

Dans cette  étape, nous avons utilisé l'\textbf{API RxNLP} pour calculer la similarité.
\vspace{-0.3cm}
\paragraph{L'API RxNLP : \\
	}{
L'API RxNLP fournit l'accès à certaines fonctionnalités avancées d'analyse de texte sur un cloud. L'API a actuellement un bon mélange d'utilisateurs commerciaux et universitaires. Pour utiliser l' API de \textbf{RxNLP} , on doit s'inscrire à un compte \textit{Mashape}\footnote{https://market.mashape.com}, puis s'abonner au plan de l'API qui convient. Cette API est hébergée sur \textit{Azure Cloud Plateforme de Microsoft}.}

Dans ce qui suit ci-dessous, une liste des opérations actuelles de la plateforme :

\vspace{0.2cm}
\begin{enumerate}
	\item \textbf{HTML2Text}
	
	Le point de terminaison \textit{HTML2Text} extraire uniquement le corps du texte d'une page HTML ou extraire le corps du texte directement à partir d'une URL.
	
	\item \textbf{N-Gram et le comptage de mots}
	
	Le point de terminaison \textit{N-Gram} et le comptage de mots; génère des comptes de mots et de n-grammes dans n'importe quelle langue. Les mots ou n-grammes et sa fréquence sont retournés dans l'ordre décroissant de la fréquence.
	
	\item \textbf{Similarité du texte}
	
	Le point de terminaison \textit{Text Similarity} calcule la similarité entre deux textes (longs ou courts), en utilisant des mesures bien connues telles que Jaccard, Dice et Cosinus.
	
	\item \textbf{Clustering de phrases (regroupement)}
	
	Le point de terminaison \textit{Clustering} de phrases regroupe des textes tels que les tweets, les tickets de support client, les articles de presse, les sondages, les revues d'utilisateurs et autres dans des groupes de phrases logiques. On obtient les clusters, le score de cluster et les étiquettes de cluster.
	
	\item \textbf{Récapitulation d'Opinosis}
	
	Le point de terminaison de la récapitulation d'\textit{Opinosis}, génère de brefs résumés d'opinions et il est conçu pour fonctionner avec des textes tels que les critiques d'utilisateurs.
	
\end{enumerate}

Nous avons utilisé cette \textbf{API} pour calculer la similarité des sujets (textes), où le choix de trois types de similarités, qui sont disponibles : \textbf{Jaccard}, \textbf{Dice} et \textbf{Cosinus} en plus nous avons aussi calculé la moyenne des trois.

\vspace{0.3cm}
Après déroulement des trois phases, des résultats ont été obtenus.

\section{Résultats obtenus}
\subsection{Résultat d'étude de similarité Utilisateur-Contenu}

L'algorithme \textit{Utilisateur-Contenu} avec les données que nous avons préparé prend : \textit{\textbf{5h 16min 49s}} en utilisant une machine de \textit{6Gb} RAM et \textit{2Mb} comme vitesse d'Internet.

\vspace{0.3cm}
\begin{pyin}	
from pyrxnlp.api.text_similarity import TextSimilarity 
apikey = "XXXXX"
t = TextSimilarity(apikey, True)
	
similarity = pd.DataFrame(index= np.arange(0,len(users)),
                          columns=['sim_cos','sim_jaccard',
                                   'sim_dice','sim_avg'])
for i in range(0,len(users)):
  sims = t.get_similarity("Business & Finance , 
                           Small business ,Internet , 
                           Technology", users.iloc[i,3])
  similarity.iloc[i,0] = sims["cosine"]
  similarity.iloc[i,1] = sims["jaccard"]
  similarity.iloc[i,2] = sims["dice"]
  similarity.iloc[i,3] = sims["average"]
	
\end{pyin}

\begin{pyout}
	time : 5h 16m 49s
\end{pyout}

\vspace{-0.5cm}
\begin{center}
	\captionof{figure}{Code Python de Calcule de similarité Utilisateur-Contenu.}
\end{center}
\vspace{-0.5cm}
Les précisions suivantes ont été obtenues :
\vspace{0.3cm}
	\begin{itemize}
		\item Avec la similarité de Cosinus = \textbf{\textit{37.1354862487 \%}}.
		\item Avec la similarité de Jaccard = \textbf{\textit{36.6609849357 \%}}.
		\item Avec la similarité de Dice = \textbf{\textit{37.1354862487 \%}}.
		\item Avec la moyenne des similarités = \textbf{\textit{37.1032385866 \%}}.
    \end{itemize}

\begin{center}
	\includegraphics[scale=0.35]{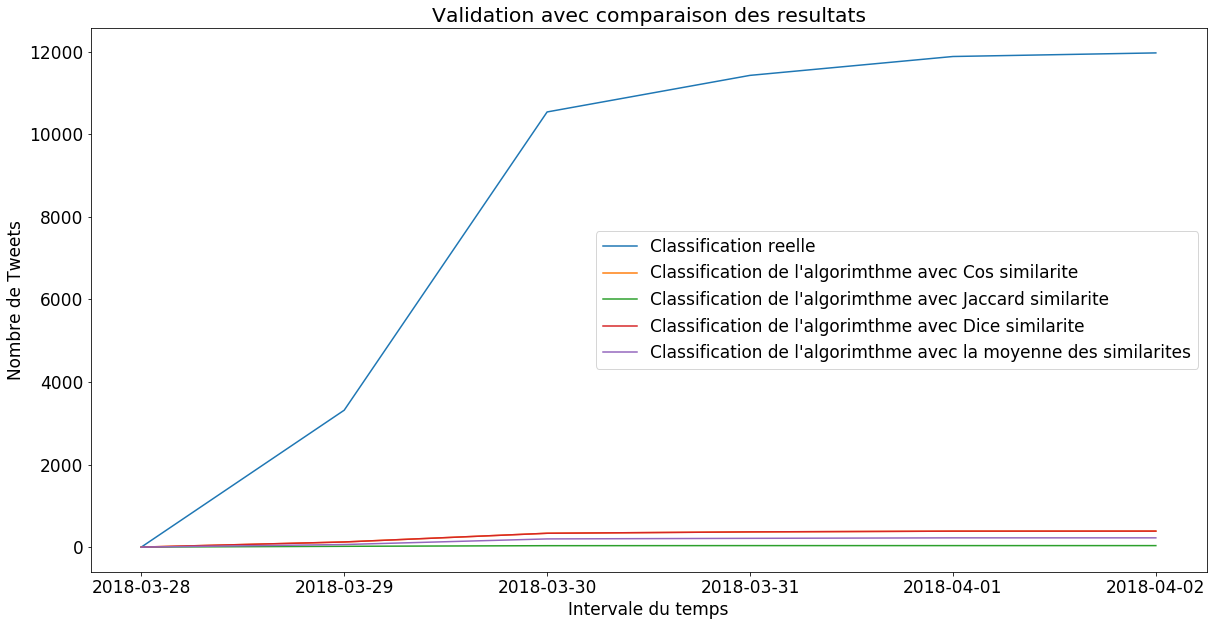}
	\captionof{figure}{Résultat d'étude de similarité Utilisateur-Contenu.}
\end{center}
\textbf{Interprétation du résultat :}

\vspace{0.3cm}
Nous avons remarqué que, la similarité d'utilisateurs avec le contenu diffusé n'a aucune relation avec la diffusion de l'information (rumeur), où nous avons obtenu un minimum d'erreur de \textit{\textbf{62.8645137513\%}}.

\subsection{Résultat d'étude de similarité Utilisateur-Utilisateur}
L'algorithme \textit{Utilisateur-Utilisateur} avec les données que nous avons préparé prend : \textit{\textbf{6h 34min 51s}} en utilisant une  machine de \textit{6Gb} RAM et \textit{2Mb} comme vitesse d'Internet.

\begin{pyin}	
from pyrxnlp.api.text_similarity import TextSimilarity 
apikey = "XXXXX"
t = TextSimilarity(apikey, True)
	
similarity = pd.DataFrame(index= np.arange(0,len(users)),
                          columns=['sim_cos','sim_jaccard',
                                   'sim_dice','sim_avg'])
for i in range(0,len(graph)):
  sims = t.get_similarity(graph.iloc[i,7], graph.iloc[i,8])
  similarity.iloc[i,0] = sims["cosine"]
  similarity.iloc[i,1] = sims["jaccard"]
  similarity.iloc[i,2] = sims["dice"]
  similarity.iloc[i,3] = sims["average"]
	
\end{pyin}

\begin{pyout}
	time : 6h 34m 51s
\end{pyout}
\vspace{-0.5cm}
\begin{center}
	\captionof{figure}{Code Python de Calcul de similarité Utilisateur-Utilisateur.}
\end{center}	
\vspace{-0.5cm}
Les précisions suivantes ont été obtenues :
\vspace{0.3cm}
\begin{itemize}
	\item Avec la similarité de Cosinus = \textbf{\textit{86.9067427304 \%}}.
	\item Avec la similarité de Jaccard = \textbf{\textit{85.6429759924 \%}}.
	\item Avec la similarité de Dice = \textbf{\textit{84.7872593297 \%}}.
	\item Avec la moyenne des similarités = \textbf{\textit{85.0923064733 \%}}.
\end{itemize}
\newpage
\begin{center}
	\includegraphics[scale=0.35]{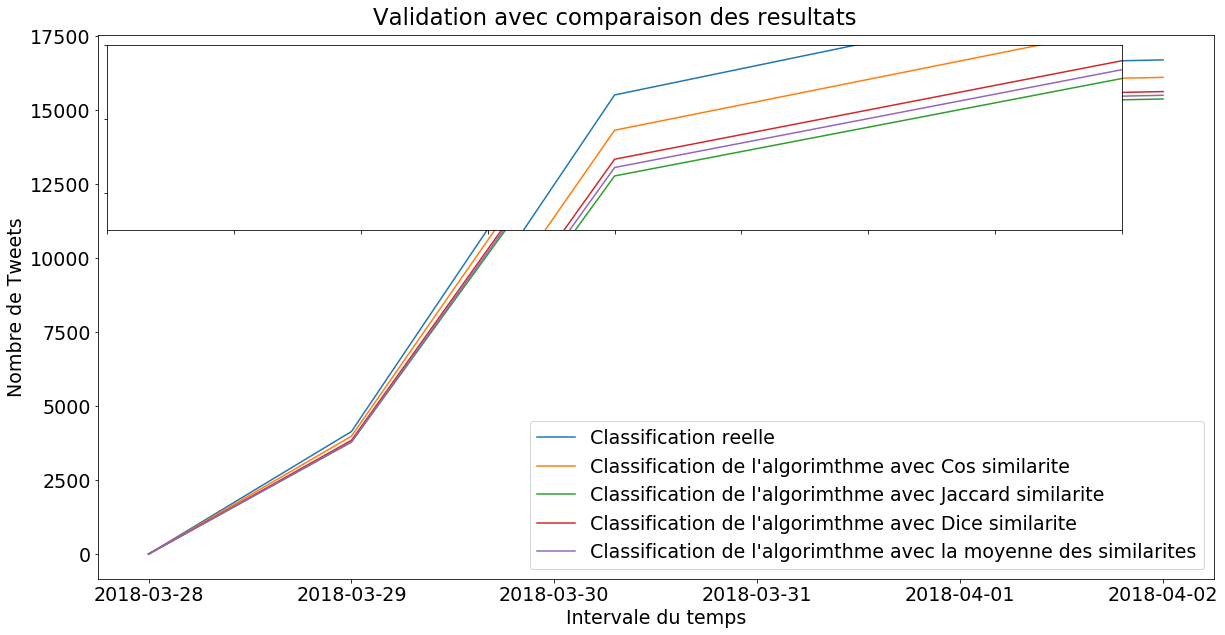}
	\captionof{figure}{Résultat d'étude de similarité Utilisateur-Utilisateur.}
\end{center}
\textbf{Interprétation du résultat :}
\begin{itemize}
	\item Nous avons remarqué que, les résultats sont très proches aux données réelles avec un minimum d'erreur de \textit{\textbf{13.0932572696\%}}, où nous pouvons déduire que la similarité entre les utilisateurs est très importante pour la diffusion de l'information (rumeur).
	\item  Nous avons remarqué aussi que, la vitesse de la diffusion a été très rapide dans les trois premiers jours, et après, la diffusion a été un peu lente (vers une stabilisation).
	\item la similarité \textbf{Cosinus} (Orange) est celle qui donne le meilleur résultat.
\end{itemize}

\section{Implémentation de l'algorithme avec la similarité Cosinus entre utilisateurs}

Depuis les résultats obtenus, nous avons implémenté l'algorithme de la diffusion en utilisant la similarité \textbf{Utilisateur-Utilisateur} avec la métrique \textbf{Cosinus}.

\begin{pyin}	
def G_followers(n):
  gf = Graph[Graph.from_user_id == n]
  gf = gf.reset_index(drop=True)
  return gf
	
Diffuseurs = []
Diffuseurs.append(Graph.iloc[0,0]) 
for noeud in Diffuseurs:
  GF = G_followers(noeud)
  if(len(GF)>0):
    for i in range(0,len(GF)):
      if(GF.iloc[i,14] == 1): # utilisant cos similarite.
        if GF.iloc[i,2] not in Diffuseurs:
          Diffuseurs.append(GF.iloc[i,2])

print "Le nombre de diffuseurs de la rumeur est : "+
str(len(Diffuseurs))
\end{pyin}

\begin{pyout}
Le nombre de diffuseurs de la rumeur est : 15762
\end{pyout}
\vspace{-0.5cm}
\begin{center}
	\captionof{figure}{Code Python de l'algorithme de diffusion \textbf{Utilisateur-Utilisateur} utilisant la similarité \textbf{Cosinus}.}
\end{center}	
\vspace{-2cm}

\section{Visualisation du graphe de diffusion}

Pour visualiser le graphe de diffusion des rumeurs, nous avons utilisé le package \textit{NetworkX} :

\vspace{0.3cm}
\begin{itemize}
	\item \textbf{NetworkX} :
	 est un package \textit{Python} pour la création, la manipulation et l'étude de la structure, de la dynamique et des fonctions de réseaux complexes.
	 
	 \item \textbf{networkx.drawing.nx\_agraph} :
	  est un fichier intégré dans le package \textit{NetworkX}. Il est utilisé pour la visualisation des grands réseaux complexes.
	  
\end{itemize}

\begin{center}
\includegraphics[scale=0.70]{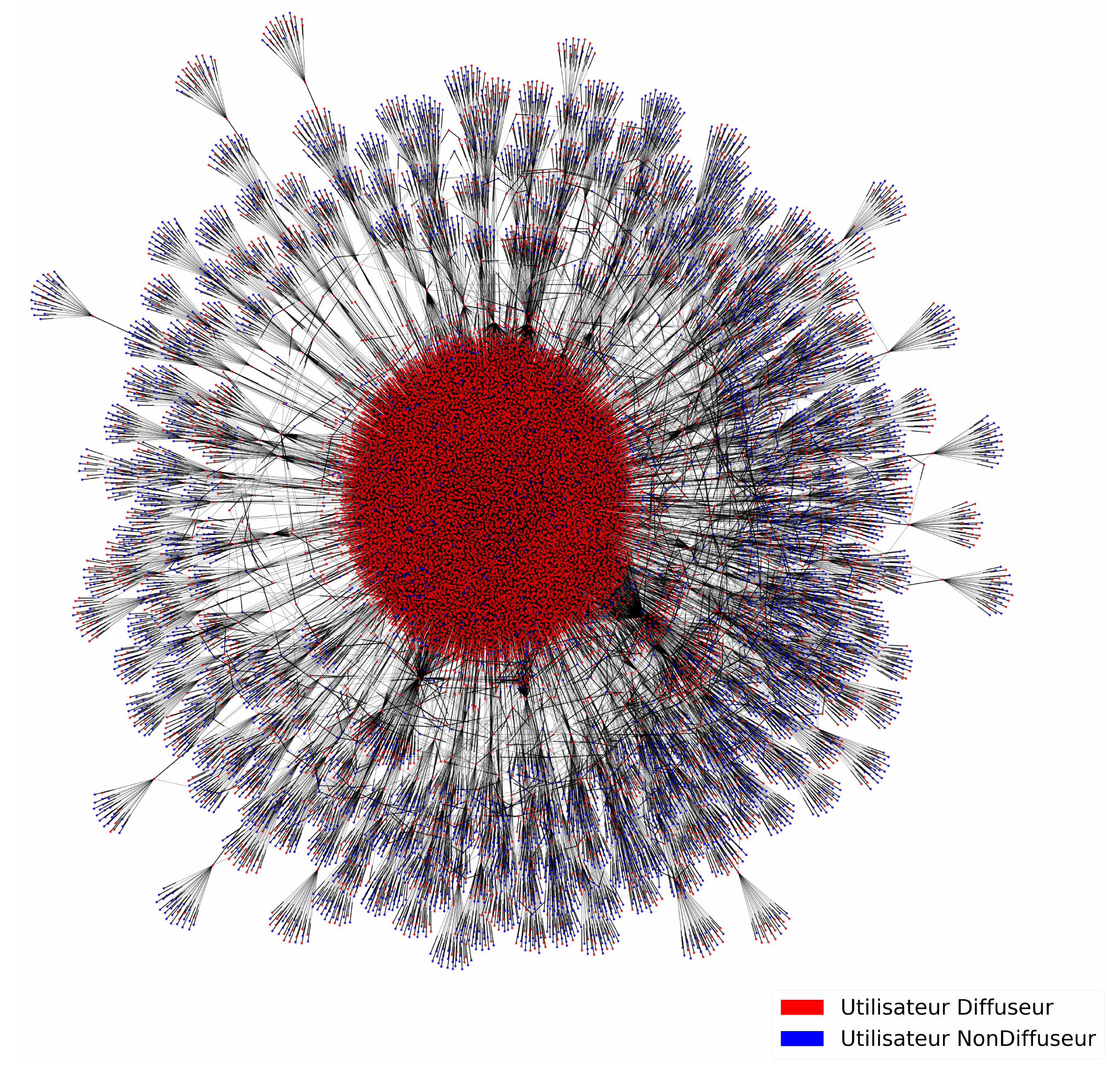}
	
	\captionof{figure}{Diffusion des rumeurs sur le réseau social \textbf{Twitter}.}
\end{center}

\subsection{Interprétation du résultat:}
\begin{itemize}
	\item Nous avons remarqué que, le plus nombreux de diffuseurs représente une communauté, où tous ces utilisateurs sont des abonnés un par un.
	
	\item Nous pouvons conclure que, le nombre de voisins diffuseurs peut affecter la décision de la diffusion.
	
\end{itemize}
\section{Simulation du processus de diffusion des rumeurs}

Pour dynamiser la visualisation de la progression de la diffusion du rumeur, nous avons simulé ce processus en utilisant la modélisation d'agents.
\subsection{Modélisation d'agent}

La modélisation basée sur l'agent est un style de modélisation dans lequel les individus et leur interaction entre eux et leur environnement sont explicitement représentés dans un programme ou même dans une autre entité physique telle qu'un robot. Les individus modélisés sont, par exemple, des personnes, des animaux, des groupes ou des cellules, mais ils peuvent modéliser des entités qui n'ont pas de base physique, mais sont conçues comme exécutant une tâche telle que la collecte d'informations ou la modélisation théorique de l'évolution de la coopération. 

\vspace{0.3cm}
Les décisions communes dans la conception de modèle basé sur l'agent sont \cite{RAND}: portée du modèle, définition des agents, propriétés des agents, comportements de l'agent, environnement, pas de temps et enfin les entrées et les sorties. Ceux-ci sont détaillés ci-dessous:
\vspace{0.3cm}
\begin{enumerate}
	\item \textit{Portée du modèle} : 
	
	C'est la partie du système cible sur laquelle le modèle est centré et quels aspects peuvent être ignorés. L'idée de développer un modèle est qu'il devrait être aussi simple que possible, afin de l'étudier facilement, mais en même temps, le modèle doit décrire la réalité. En ce qui concerne le cas \textbf{Twitter}, il y a un certain nombre de phénomènes qui peuvent être intéressant à des fins de marketing: propagation de retweets, nombre de mentions, nombre de tweets avec un Hashtag spécifique, activité par fuseau horaire,...etc.
	\newpage
	\item \textit{Agents} :
	
	 Une autre décision qui ne signifie pas nécessairement des agents intelligents capables, entre autres, d'apprendre  \cite{nwana_1996}; ou des agents délibératifs qui prennent des décisions en utilisant un raisonnement symbolique \cite{Woolridge}. L'agent typique est un agent réactif qui interagit de manière autonome avec les autres en se basant sur un modèle de comportement, qui peut varier des règles de production aux modèles d'apprentissage machine, tels que les réseaux neuronaux artificiels \cite{CAMPUZANO}. Dans le cas \textbf{Twitter}, la décision directe est d'avoir des agents pour chaque utilisateur de \textbf{Twitter}.
	 
	 \item \textit{Propriétés} :
	 
	 Ce sont les champs qui décrivent chaque agent. Encore une fois, ceux-ci dépendent complètement de la portée du modèle. Pour le cas de propagation de la rumeur Twitter, comme expliqué, les propriétés typiques incluent: un identifiant; une position dans l'environnement ; l'état de l'agent par rapport à la rumeur (Diffuseur,Non-Diffuseur,...etc.); et, si nécessaire, et un champ de type agent qui peut déterminer l'étendue d'autres propriétés ou le comportement de l'agent.
	 
	 \item \textit{Comportements} :
	 
	 Les agents présentent un comportement qui implique l'interaction avec l'environnement et d'autres agents à chaque étape. Généralement, ces comportements sont des processus stochastiques qui dépendent de probabilités données. Il existe d'innombrables manières de définir les comportements : les règles de production \cite{SerranoCG14}, les modèles d'apprentissage automatique \cite{Serrano2013}, fonctions de densité de probabilité \cite{GarciaValverde},...etc. Ainsi, les auteurs recommandent l'utilisation de pseudo-code ou toute autre technique de modélisation logicielle générale pour les logiciels généraux tels que les diagrammes \textit{UML}. L'utilisation de diagramme de \textit{flux}, qui correspondent grosso modo aux diagrammes d'activités \textit{UML}, est très populaire dans la littérature \cite{Gilbert}.
	 \newpage
	 \item \textit{Environnement} :
	 
	 L'environnement définit la topologie d'interaction des agents. Pour \textit{Twitter}, l'environnement est largement décrit comme un réseau ou un graphique où les nœuds représentent les utilisateurs. D'autres types de réseaux sont également possibles en fonction de la portée, tels que les réseaux de \textit{retweets}. Les liens n'ont pas la même signification dans tous les travaux. Alors que certains auteurs représentent l'asymétrie de \textit{Twitter} (utilisateur $ u_{2} $ suit $ u_{1} $), $ u_{1} $ suit $ u_{2} $ ne veut pas dire que les autres considèrent les liens indirects puisque \textit{Twitter} a des mécanismes pour faire circuler l'information du suiveur au suivant (comme les réponses, les mentions, les retweets et les messages privés).
	 
	 \item \textit{Pas de temps} :
	 
	 Les Agents évoluent généralement dans le temps en utilisant un \textit{pas de temps}. Deux phases sont distinguées: l'initialisation, lorsque les agents et l'environnement sont créés; et itération, où les agents agissent en fonction de leur modèle de comportement. De plus, en fonction de la portée et des données réelles disponibles, le pas de temps représentera une unité de temps physique différente. Pour le cas de \textbf{Twitter}, les tâches d'analyse des données et d'analyse exploratoire des données peuvent donner un aperçu de cette décision. Si les données sont très dispersées (par exemple, les jours passent entre les tweets pertinents), il n'y a pas suffisamment d'informations pour un pas de temps court (par exemple, la simulation d'heures).
	 
	 \item \textit{Entrée et sortie} :
	 
	 Les paramètres et les valeurs observés dans l'exécution de la simulation sont d'autres décisions pour le modèle. L'une des entrées les plus importantes et couramment utilisées dans l'agent est la graine aléatoire. Comme on le voit, un agent implique un certain nombre de processus stochastiques: sélection de l'ordre d'exécution des agents à un pas de temps, création d'un modèle de réseau, décision parmi les actions possibles dans le modèle de comportement,...etc.
	 
	 Un défaut majeur de la recherche agent, n'est pas de s'assurer que tous ces processus dépendent d'une seule graine aléatoire, perdant la répétabilité et la reproductibilité de la simulation.
	 
\end{enumerate}


\subsection{Implémentation de la simulation}

 Pour implémenter la simulation sous le langage de programmation \textbf{Python}, nous avons utilisé \textbf{ComplexNetworkSim} qui est un package \textit{Python} pour la simulation d'agents connectés dans un réseau complexe, mais avec une personnalisation dans le fichier de la représentation graphique [Annexe A], où nous avons obtenu le modèle suivant :
 
 \vspace{0.3cm}
 
\begin{itemize}
	\item L'\textit{utilisateur} représente un \textbf{Agent}.
	\item \textit{Suit (Follow) } représente la relation entre un \textbf{Agent} et ses \textbf{Agents voisins}.
	\item Chaque \textit{Agent} a un comportement.
		
\end{itemize}

\begin{pyin}	
class User(NetworkAgent):
  def __init__(self, state, initialiser):
    NetworkAgent.__init__(self, state, initialiser)
    self.sim_end = 3000
    self.Created_at = Graph["Created_at"][self.id]
	
  def Run(self):
    while True:
      if self.state == NonDiffuseur:
        yield Sim.hold,self, self.Created_at
        self.peutEtreDiffuseur() 
      elif self.state == Diffuseur:
        yield Sim.hold, self, self.sim_end
	
  def peutEtreDiffuseur(self):
    voisins_diffuseurs = self.getNeighbouringAgents(state =
                                                 Diffuseur)
    for voisin in voisins_diffuseurs:
     if((Graph["pred_cos"][(Graph.from_user_id ==voisin.id)
        &(Graph.to_user_id == self.id)]).empty == False):
      if((Graph["pred_cos"][(Graph.from_user_id==voisin.id) 
         &(Graph.to_user_id == self.id)] == 1).bool()):
         self.state = Diffuseur
	     break
\end{pyin}
\begin{center}
	\captionof{figure}{Comportement de l'Agent (Behaviour).}
\end{center}
\vspace{-1cm}
Chaque Agent, durant la simulation, a un comportement avec une initialisation des différents paramètres (l'état initial, le temps d'exécution,...etc) et aussi un fonctionnement à exécuter. Dans le cas de notre agent \textbf{User} : 
\vspace{-0.2cm}
\begin{Verbatim}
	Si l'état de l'agent est NonDiffuseur alors
		il va bloquer et attendre jusqu'au temps de son
		exécution où il va être réveillé pour exécuter
		l'algorithme de diffusion.
	Sinon (état Diffuseur) 
		il va attendre jusqu'à la fin de la simulation.   
\end{Verbatim}

\begin{pyin}	
from ComplexNetworkSim import NetworkSimulation
# Constantes de Simulation :
MAX_SIMULATION_TIME = 1296
TRIALS = 2
def main():
directory = 'pfe' # output directory
# executer la  simulation avec les parametres
# - complex network structure : G
# - liste initiale des etats : states
# - agent behaviour classe : User
# - dossier de sortie : directory
# - temps maximum de simulation : MAX_SIMULATION_TIME
# - nombre de boucle : TRIALS
simulation = NetworkSimulation(G,states,User,directory,
                               MAX_SIMULATION_TIME,TRIALS)
simulation.runSimulation()
if __name__ == '__main__':
main()

\end{pyin}
\vspace{-0.5cm}
\begin{center}
	\captionof{figure}{Processus de simulation.}
\end{center}	
\vspace{-0.6cm}
Avant de pouvoir manipuler une simulation sous le package \textbf{ComplexNetworkSim}, il faut d'abord spécifier un ensemble de paramètres qui sont les suivants :

\vspace{0.3cm}
\begin{itemize}
	\item \textbf{La structure du réseau :} un graphe \textit{G}.
	\item \textbf{La liste initiale des états des agents:}  une liste (\textit{states}).
	\item \textbf{La classe de l'Agent :} la classe \textit{User}.
	\item \textbf{Un répertoire de sortie :} où les résultats sont enregistrés (\textit{pfe}). 
	
	\item \textbf{Le temps maximum de simulation :} combien de temps nous voulons simuler (\textit{MAX\_SIMULATION\_TIME}).
	\item \textbf{Le nombre d'essais de simulation :}l'idée est que nous pouvons simuler un scénario plusieurs fois avec les mêmes entrées puis prendre une moyenne sur ce qui se passe, si cela est pertinent dans notre cas particulier (\textit{TRIALS}).

\end{itemize}

Ensuite, nous créons une instance de \textbf{NetworkSimulation} avec nos paramètres et appelons sa méthode \textit{.runSimulation ()}.

\begin{pyin}	
from ComplexNetworkSim import PlotCreator, AnimationCreator

directory = 'pfe' # output directory 
myName = "Rumor diffusion" # nom de output 
title = "Simulation de diffusion des rumeur sur Twitter"
statesToMonitor = [Diffuseur, NonDiffuseur] 
colours = ["r", "b"] 
labels = ["Diffuseur", "NonDiffuseur"]
mapping = {NonDiffuseur:"b", Diffuseur:"r"}
trialToVisualise = 0

p = PlotCreator(directory, myName, title, statesToMonitor,
                colours, labels)
p.plotSimulation(show=True)

visualiser = AnimationCreator(directory, myName, title, 
                           mapping, trial=trialToVisualise)
visualiser.create_gif(verbose=True)	
\end{pyin}
\begin{center}
	\captionof{figure}{Processus de visualisation de simulation.}
\end{center}

\vspace{-0.5cm}
Après avoir créé la simulation, et vérifié sa fonctionnalité et sa production des fichiers de sortie \textit{pickled}, maintenant, nous souhaitons visualiser les résultats. 

\vspace{0.3cm}
Le package \textit{ComplexNetworkSim} prend en charge la génération de graphiques de base et la visualisation des changements d'état du réseau au fil du temps. Nous avons chargé d'abord les classes: \textbf{PlotCreator} et \textbf{AnimationCreator} Ensuite, nous avons défini quelques paramètres tels que : les noms, les états et les couleurs pour les nœuds animés en fonction de leur état.

\vspace{0.3cm}
Pour l'appel réel au traçage et à l'animation, nous créons simplement une instance de la classe \textbf{PlotCreator} avec les paramètres définis et appelons sa méthode \textit{.plotSimulation ()}. De même, pour la visualisation; nous créons une instance de la classe \textbf{AnimationCreator} et nous appelons \textit{.create\_gif()}. 

\newpage
En résultat graphique, nous avons obtenu  des fichiers image \textbf{PNG} pour chaque pas de temps plus un \textbf{gif} animé de la simulation de taille \textit{\textbf{554.4 MB}}, où nous avons pu le convertir à un fichier \textbf{MP4} de taille \textbf{\textit{3.7 MB}} de durée de \textbf{\textit{5.25 }} minutes pour pouvoir le lire.

\subsection{Résultats de simulation}

Pour mieux visualiser la simulation, comme il est impossible de visualiser la totalité des liens que nous avons utilisé (\textbf{\textit{25242}} liens), nous avons pris un sous-ensemble de graphe de \textit{\textbf{3000}} liens, et appliquer le processus de simulation, où nous avons pu obtenir les résultats suivants :

\begin{center}
	\includegraphics[scale=0.5]{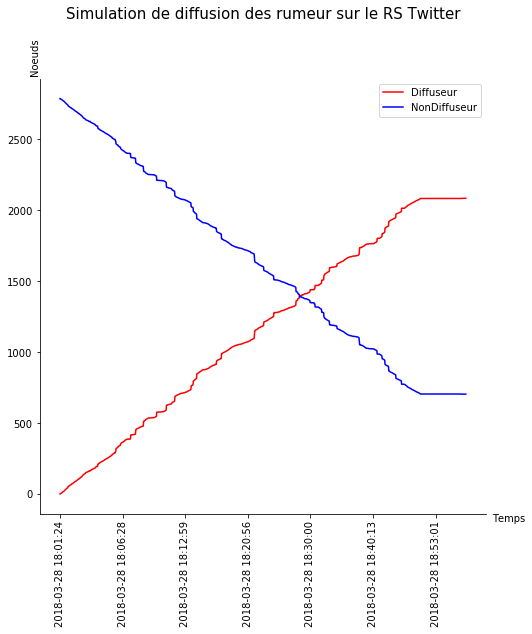}
	\captionof{figure}{Progression de nombre de diffuseurs durant la simulation du processus de diffusion des rumeurs.}
\end{center}
\begin{center}
	\includegraphics[scale=0.35]{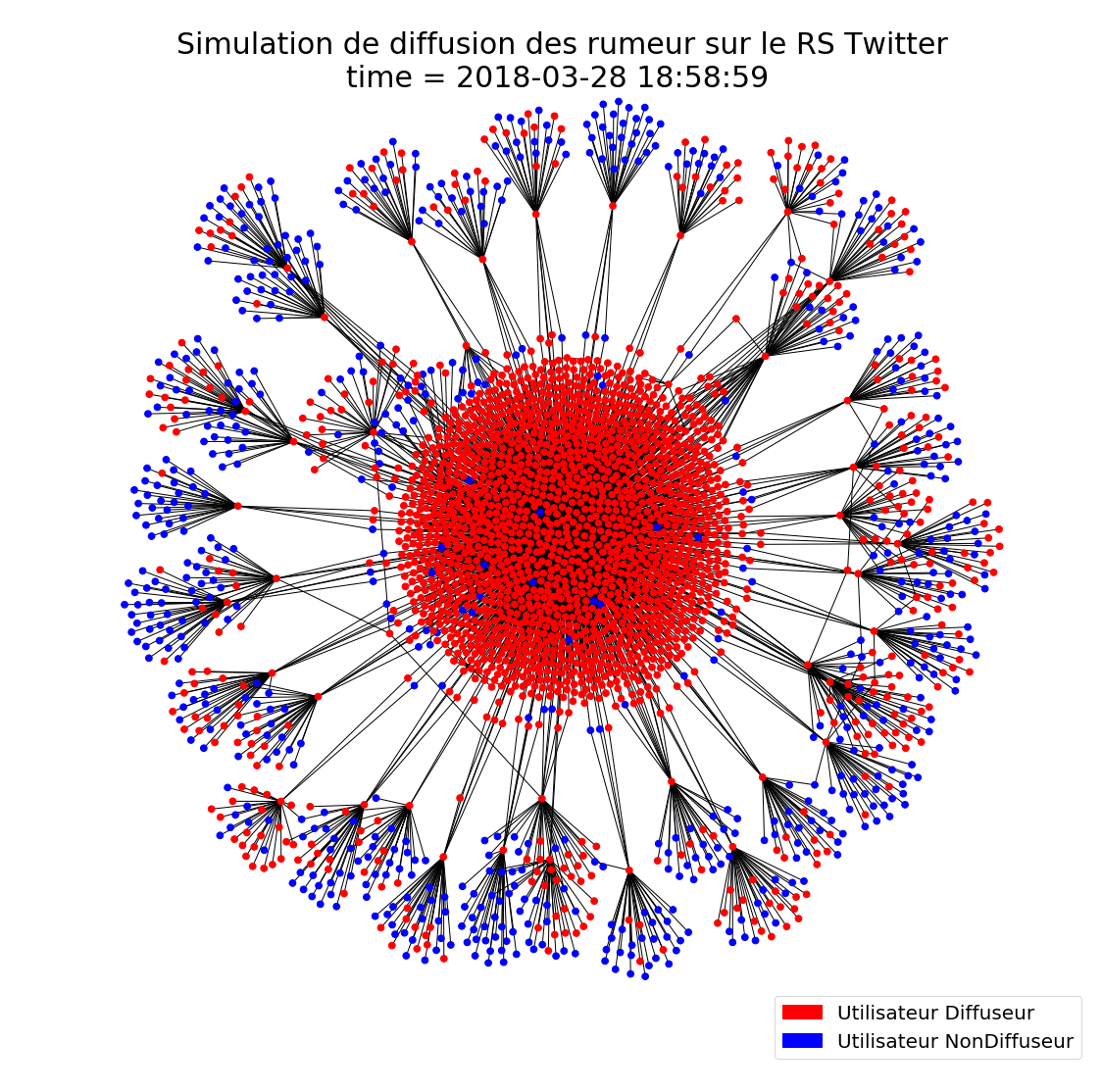}
	\captionof{figure}{Visualisation du résultat final de simulation du processus de diffusion des rumeurs.}
\end{center}

\subsection*{Interprétation des résultats :}
En principe, une meilleure interprétation des résultats dépend de la nature de la rumeur. La quantification de la profondeur d'une rumeur varie d'un cas à un autre. Les spécialistes en sciences humaines peuvent se baser sur nos résultats pour avoir une interprétation fiable et avoir l'impact de cette rumeur sur les internautes et en général sur la société.

\vspace{0.3cm}
Pour notre cas présent, nous remarquons que la rumeur va se stabiliser après un certain temps et cela peut être différent d'un cas à un autre, c-à-d que le temps que la rumeur est confirmée et contredite par un moyen officiel.

 \vspace{5cm}
\section{Conclusion}

Dans ce chapitre, nous avons présenté une étude de cas sur le réseau social Twitter, où la réalisation d'un ensemble de phases pour évaluer les performances de nos algorithmes proposés sur une base de données réelles.

\vspace{0.3cm}
D'abord, nous avons collecté et traité les données en utilisant un ensemble d'\textbf{APIs}, puis appliqué notre approche sur ces données réelles. Nous avons obtenu comme résultats que la diffusion des rumeurs sur les réseaux sociaux ne dépend pas de la similarité Utilisateur-Contenu mais de la similarité Utilisateur-Utilisateur.
Aussi, en terme de contenu de sujets, la meilleure métrique de similarité qui donne le meilleur résultat est la similarité \textbf{Cosinus}.

%% file: Chapitres/Conclusion.tex
\chapter*{Conclusion générale}
\markboth{\textit{Conclusion générale}}{}
\addcontentsline{toc}{chapter}{Conclusion générale}

Dans ce présent travail, nous avons abordé le problème de la quantification de la profondeur de la diffusion d'une rumeur sur les réseaux sociaux. Contrairement aux approches traditionnelles qui se concentrent sur différents types de modèles de diffusion pour comprendre ces phénomènes, nous avons abordé ici le problème d'un autre point de vue, en étudiant la diffusion à partir d'une situation réelle survenant sur le site de micro-blogging \textbf{Twitter}.

\vspace{0.3cm}
En effet, les réseaux sociaux sont devenus un véritable \textit{"phénomène"} de société, car ils ont largement modifié la manière dont nous produisons, diffusons et consommons l’information et fait de chacun une source d'information. La rumeur est émise et répétée plus rapidement et plus largement que jamais, en raison de la connectivité des réseaux sociaux, qui ont un impact sur le monde réel. Un faux tweet a un impact négatif sur la communauté ou la famille, déclenche une panique financière et met même à rude épreuve les relations diplomatiques.

\vspace{0.3cm}
Dans ce cadre, nous avons entamé ce travail par une prospection des réseaux sociaux et leurs relations avec la propagation des rumeurs. Aussi, un passage par un état de l'art sur ce qu'a été déjà fait dans ce sens, a été bénéfique pour nous. Comme le phénomène de propagation des rumeurs sur les réseaux sociaux est très récent, l'apport académique est aussi très limité, mais nous avons pu exploiter un ensemble de travaux existant sur la modélisation de diffusion d'information généralement et spécialement de la rumeur, que nous avons jugés utiles pour notre contribution, où nous avons proposé une nouvelle étude basée sur la notion de similarité; afin de calculer la probabilité de diffuser un contenu donné, en étudiant l'effet de la similarité utilisateur-utilisateur et contenu-utilisateur sur la propagation de la rumeur basé sur les sujets d'historique de partage de chaque utilisateur, en utilisant différentes métriques de similarités de textes les plus connues (Cosinus, corrélation de Pearson, Jaccard, Levenshtein et Dice).

\vspace{0.3cm}
En guise, d'obtenir des résultats fiables, nous avons utilisé des données réelles à partir du réseau social \textbf{Twitter}, en suivant un ensemble de phases.
 
\vspace{0.2cm}
Premièrement, nous avons re-utilisé deux types d'API afin de télécharger les données utilisateurs. Et nous avons pu collecter \textbf{\textit{17083}} utilisateurs avec \textbf{\textit{25242}} liens entre eux. 

\vspace{0.2cm}
Deuxièmement, nous avons appliqué une extraction des sujets des tweets, en utilisant l'\textbf{API MonkeyLearn}.

\vspace{0.2cm}
Puis l'\textbf{API RxNLP} pour calculer la similarité, en utilisant trois métriques (Cosinus, Jaccard et Dice).

\vspace{0.3cm}
Finalement, Nous avons évalué nos algorithmes en les déroulant sur ces données et les résultats sont très importants. 

\vspace{0.2cm}
Premièrement, nous avons trouvé que la similarité d'utilisateurs avec le contenu diffusé, n'a aucune relation avec la diffusion de l'information (rumeur), avec l'obtention d'un minimum d'erreur de \textbf{\textit{62.86\%}}. 

\vspace{0.2cm}  
Deuxièmement, nous pouvons déduire que la similarité entre les utilisateurs est très importante, pour la diffusion de l'information (rumeur), car les résultats sont très proches aux données réelles avec un minimum d'erreur de \textbf{\textit{13.09\%}}; c.à.d une précision de \textbf{\textit{86.90\%}}, en utilisant la similarité \textbf{Cosinus} qui a donné le meilleur résultat.

\vspace{0.3cm}
Enfin, nous considérons que ce travail ouvre également des perspectives intéressantes en termes de modélisation réalistes, tenant compte des comportements individuels et la sémantique du phénomène qui pourraient être en mesure de prédire, pour un utilisateur donné, s’il transmettra une information ou pas.

%% file: Chapitres/Annexe.tex
\begin{appendix}
\chapter{Annexe : Personnalisation de package ComplexNetworkSim}
	\label{annexeA}
    \section*{Le fichier "animation.py" :}
	\subsection*{La fonction "createPNGs" :}
	
\begin{pyin}
def createPNGs(self):      
  states, topos, vector = utils.retrieveTrial(self.dir
                                              ,self.trial)
  init_topo = topos[0][1]
  if topos[0][0] != 0:
    print "problem - first topology not starting at 0!"
  self.G = init_topo
  # Chagement concernant le Layout :
  self.layout = graphviz_layout(self.G, prog='sfdp')
  self.nodesToDraw = self.G.nodes()
  self.edgesToDraw = self.G.edges()
  self.nodesToDraw.sort()
  self.edgesToDraw.sort()
  i = 1 
  j = 0
  # Telechargement de donnees temporelles :
  G = pd.read_csv(os.path.join(self.dir,"time.csv"))
\end{pyin}
\begin{pyin} 
  .# Changement de resolution des images sortantes :
  pyplot.figure(figsize=(15,15))
  for t, s in states:
    if len(topos) > i and t == topos[i][0]:                
      self.G = topos[i][1]
      fixednodes = [node for node in self.nodesToDraw 
                    if node in self.G.nodes()]                
      self.layout = graphviz_layout(self.G, prog='sfdp')
      i += 1
    colours = utils.states_to_colourString(s, self.mapping)            
    pyplot.clf()
    self.nodesToDraw = self.G.nodes()
    self.nodesToDraw.sort()
    self.edgesToDraw = self.G.edges()
    self.edgesToDraw.sort()
    # utilisation de Networkx pour la visualisation :
    nx.draw(self.G, self.layout , with_labels=False,
            arrows=True,node_size=45,node_color=colours,
            alpha=1,width = 1)
    # Visualisation de legende et titres :
    red_patch = mpatches.Patch(color='red', 
                             label='Utilisateur Diffuseur')
    blue_patch = mpatches.Patch(color='blue',
                           label='Utilisateur NonDiffuseur')
    ti = G["date"][G.num == t]
    pyplot.suptitle("\%s\ntime = \%s \n" \% (self.title,
                    ti.iloc[0]),size = 30)
    pyplot.legend(handles=[red_patch,blue_patch],
                 prop={'size': 20},loc=4)
    pyplot.axis('off')
    import cStringIO
    ram = cStringIO.StringIO()
    pyplot.savefig(ram, format='png')
    ram.seek(0)
    im = Image.open(ram)
    im2 = im.convert('RGB').convert('P', palette=(8, 8))
    im2.save( os.path.join(self.dir, 
             self.name +" " +"\%02d.png" \% j))
    j = j+1
	
\end{pyin}

\end{appendix}

%% file: Chapitres/Resume.tex
\thispagestyle{empty}
\begin{small}
\section*{\abstractname}
\vspace{-0.2cm}
La propagation d'une rumeur (information \textit{non-vérifiée}) sur un réseau social est assujettie  à plusieurs facteurs liés essentiellement au contenu de cette information et surtout aux comportements (profils) des acteurs sur ce réseau qui la retransmettent. Cet état de fait peut faire varier cette propagation selon le cas, et, c'est ce que nous appelons profondeur de la rumeur. Ce projet s'attaque justement à cette problématique. A partir d'un cas réel de la propagation d'une rumeur sur Twitter, cette contribution propose une démarche académique afin de quantifier la profondeur d'une rumeur sur réseau social et ce, pour une utilisation et interprétation, par des spécialistes concernés par la nature de cette information et son auditorat.

\vspace{0.2cm}
\textbf{Mots-clé :} Réseaux sociaux, propagation de rumeurs.

\begin{otherlanguage}{english}
\vspace{-0.7cm}
\section*{\abstractname}
\vspace{-0.2cm}
The propagation of a rumor (\textit{unverified} information) on a social network is subject to several factors mainly related to the content of this information and especially to the behaviors (profiles) of the actors on this network that retransmit. This state of affairs may vary this propagation as the case may be, and this is what we call the depth of the rumor. This project is tackling this problem. From a real case of the spread of a rumor on Twitter, this contribution proposes an academic approach to quantify the depth of a rumor on social network and this, for a use and interpretation, by specialists concerned by the nature of this information and its auditor.

\vspace{0.2cm}
\textbf{Keywords :} Social Networks, Rumor Spreading.
\end{otherlanguage}

\thispagestyle{empty}
\begin{otherlanguage}{arabic}
\vspace{-0.7cm}
\section*{ملخص}
\vspace{-0.2cm}
انتشار الإشاعة  (المعلومة الغير المؤكدة) على شبكة التواصل الاجتماعية،  يخضع لعدة عوامل تتعلق بشكل أساسي بمحتوى هذه المعلومات وخاصةً بسلوكيات  مستعملي هذه الشبكة (الملفات الشخصية) الذين  يعيدون نشرها. و هذا الانتشار يختلف على حسب الحالة ، وهو ما يسمى بعمق الإشاعة. و من هدف هذا المشروع ، معالجة هذه المشكلة. انطلاقا من واقع انتشار إشاعة على تويتر ، نقترح مقاربة أكاديمية في  هذه المساهمة لتقدير عمق انتشار إشاعة على شبكة تواصل اجتماعية وهذا ، من أجل الاستخدام والتفسير ، من قبل المتخصصين المعنيين بطبيعة هذه المعلومات و جمهورها.

\vspace{0.2cm}
\textbf{الكلمات الدالة :} الشبكات الاجتماعية ، انتشار الشائعات.

\end{otherlanguage}
\end{small}
\cleardoublepage